\documentstyle[12pt]{article}

\makeatletter

\@addtoreset{equation}{section}
\def\section{\@startsection {section}{1}{\z@}{-3.5ex plus -1ex minus
 -.2ex}{2.3ex plus .2ex}{\large\bf}}
\def\subsection{\@startsection{subsection}{2}{\z@}{-3.25ex plus%
 -1ex minus -.2ex}{1.5ex plus .2ex}{\sc}}

\advance \voffset by -1.0in
\advance \hoffset by -0.6in
\def\tr{\mbox{tr}}
\def\cd{\!\cdot\!}

\def\bea{\begin{eqnarray}}
\def\eea{\end{eqnarray}}
\def\bpi{{\mbox{\boldmath $\pi$}}}

\def\btau{{\mbox{\boldmath $\tau$}}}

\def\bom{{\mbox{\boldmath $\omega$}}}
\def\bOm{{\mbox{\boldmath $\Omega$}}}

\def\bone{{\mbox{\bf 1}}_2}

\def\phit{{\tilde \phi}}
\def\thetat{{\tilde \theta}}

\def\be{{\mbox{\boldmath $e$}}}

\def\bs{{\mbox{\boldmath $s$}}}
\def\bt{{\mbox{\boldmath $t$}}}

\def\bn{{\mbox{\boldmath $n$}}}

\def\bx{{\mbox{\boldmath $x$}}}
\def\by{{\mbox{\boldmath $y$}}}

\def\bu{{\mbox{\boldmath $u$}}}

\def\bR{{\mbox{\boldmath $R$}}}
\def\bRh{\hat{\mbox{\boldmath $R$}}}
\def\bS{{\mbox{\boldmath $S$}}}

\def\bI{{\mbox{\boldmath $I$}}}
\def\bJ{{\mbox{\boldmath $J$}}}
\def\bK{{\mbox{\boldmath $K$}}}

\def\bL{{\mbox{\boldmath $L$}}}
\def\b0{\bf 0}

\def\d{\dagger}
\def\Uh{\hat U}
\def\Rh{\hat R}

\def\Tatt{T_{\mbox{\tiny att}}}
\def\frac#1#2{{\textstyle{#1\over #2}}}

\def\ba{{\mbox{\boldmath $a$}}}
\def\bu{{\mbox{\boldmath $u$}}}

\def\ba{{\mbox{\boldmath $a$}}}
\def\bb{{\mbox{\boldmath $b$}}}
\def\boldf{{\mbox{\boldmath $f$}}}
\def\bF{{\mbox{\boldmath $F$}}}
\def\balpha{{\mbox{\boldmath $\alpha$}}}
\def\bbeta{{\mbox{\boldmath $\beta$}}}

\def\tuu#1{[\tau_{#1},\Uh]\Uh^{\d}}

\begin{document}
\baselineskip 18pt
\parskip 6pt
\begin{flushright}
 12 December  1994

 DTP 94-47\\
 NI 94037\\
hep-ph/9502405

\end{flushright}

\vspace {0.5cm}

\begin{center}

{\Large \bf Attractive Channel Skyrmions and the Deuteron}

\vspace{0.8cm}

{\large

{\bf R.A. Leese}\\
Mathematical Institute\\
24-29 St. Giles\\
Oxford OX1 3LB, UK\\
e-mail: leese@maths.ox.ac.uk\\

\vspace{0.4cm}

{\bf N.S. Manton}\\
 Department of Applied Mathematics and Theoretical Physics\\
Silver Street \\
Cambridge CB3 9EW, UK\\
e-mail: nsm10@amtp.cam.ac.uk\\

\vspace{0.4cm}

{\bf B.J. Schroers}\\
 Department of Mathematical Sciences\\
South Road\\
Durham DH1 3LE, UK \\
e-mail: b.j.schroers@durham.ac.uk\\}

\vspace{1.0cm}

{\it to appear in Nuclear Physics B}

\vspace{0.5cm}

{\bf Abstract}

\end{center}

\baselineskip 12 pt

{\small
\noindent
The deuteron is described as a quantum state on a ten-dimensional
manifold $M_{10}$ of Skyrme fields of degree two, which are obtained
by calculating the holonomy of $SU(2)$ instantons. The manifold $M_{10}$
includes both  toroidal configurations of minimal energy and
configurations which are approximately the product of two Skyrmions
in the most attractive relative orientation. The quantum Hamiltonian
is of the form $-\Delta +V$, where $\Delta$ is the covariant Laplace
operator on $M_{10}$ and $V$ is the potential  which $M_{10}$ inherits
from the Skyrme  potential energy functional. Quantum states are
complex-valued functions on the double cover of $M_{10}$
satisfying certain constraints.  There is a unique bound state with the
quantum numbers of the deuteron, and its binding energy is approximately
6 MeV. Some of the deuteron's electrostatic and magnetostatic properties
are also calculated and compared with experiment.  }

\baselineskip 18pt

\pagebreak

\section{Introduction}

A fundamental challenge in particle physics is  to understand nuclear
forces from first principles. So far, there is no quantitative
understanding
of nuclear binding starting with quarks and QCD, but one may attempt
to investigate nuclei in terms of an effective low energy theory, like the
Skyrme model \cite{Skyrme}.
In the Skyrme model, nucleons are solitons,
 and the parameters of the model are
fixed so that the masses of the nucleon and the delta resonance
are in agreement with experiment. All of nuclear physics can in principle
then be derived from the Skyrme model (assuming that it is at  least
approximately right).

Further simplifications are necessary to make progress with this programme.
It is generally agreed that treating the Skyrme model as a quantum field
theory is very hard, and  to study nucleons and their interactions  one
needs
to reduce the Skyrme model to a finite-dimensional quantum mechanics.
This can be done by picking out, as naturally as possible, a
$6N$-dimensional
set of Skyrme fields to model $N$ nucleons. One nucleon is a quantised
state
of a Skyrmion, the lowest energy classical configuration with unit baryon
number. The Skyrmion has six classical degrees of  freedom -  three
translational
and three rotational -  and their quantisation gives a nucleon with
momentum
and correlated spin and isospin. There are various suggestions for a
twelve-dimensional
 set $M_{12}$ of
two-Skyrmion fields. All include the product of two well-separated single
Skyrmions (as suggested by Skyrme in \cite{Skyrme2}) to model
well-separated nucleons,
 but it
has been known for years that the product ansatz is not good for
Skyrmions close together. The minimal energy two-Skyrmion fields are
 configurations with  toroidal  symmetry in which two separated Skyrmions
have coalesced. These should be included in $M_{12}$.

The idea we favour, in principle, is to define $M_{12}$
 as the unstable manifold
of the hedgehog two-Skyrmion configuration. This is discussed in detail in
\cite{M1}.
 There is a quantum Hamiltonian for the motion on $M_{12}$,
of the form
$H = -\Delta + V$, where $\Delta$ is the Laplacian on $M_{12}$ and $V$ is the
potential energy. In this paper we  do not consider this Hamiltonian, but a
further simplification. It is known that $M_{12}$ has two naturally
defined,
low-lying submanifolds, which should be most important for low-energy
physics
of two nucleons. The first, denoted $M_8$, is eight-dimensional and  consists
of the toroidal
configurations of minimal energy.  Braaten and Carson  considered  the
quantum
mechanics on $M_8$ in \cite{BC}. They calculated the Hamiltonian  and found
the lowest energy stationary state compatible with the nucleons being
treated
as fermions. This state is identified with  the deuteron. Not surprisingly,
Braaten and Carson found a very large binding energy, and a small physical
size. This is because the Skyrmions are not allowed to explore the full
twelve-dimensional
$M_{12}$, but only the eight-dimensional $M_8$. Nevertheless,
Braaten
and Carson's analysis is very instructive and is an important inspiration
for our
work.

We have considered the quantisation on a ten-dimensional submanifold
$M_{10}$
of  $M_{12}$. Again the Hamiltonian is derived from the kinetic and
potential parts
of the Skyrme model Lagrangian, restricted  to the fields in $M_{10}$.
The manifold $M_{10}$ is the set of attractive channel fields,
which is a low-lying
valley
in $M_{12}$.
It can be defined  by taking two well-separated  Skyrmions,
oriented so that they maximally attract (i.e. one is rotated by
$\pi$ relative
to the other, with the axis of rotation orthogonal to the line joining
them),
and allowing them to approach until they coalesce into a toroidal
configuration.
Ideally one would follow a gradient flow curve, or a path of steepest
descent.
The ten coordinates of $M_{10}$ are accounted for by the separation
parameter,
overall translations, rotations and iso-rotations. $M_{10}$ should
be thought of as the smallest subset of $M_{12}$ which includes Skyrmion
separation.

In practice, we have not followed the prescription above for constructing
$M_{10}$. Instead, we have defined  an approximation to  $M_{10}$
consisting
of Skyrme fields generated by Yang-Mills instantons. The details
of the construction are explained in section 3. The instanton approach
 has some advantages which compensate for the fact that the fields do not
have quite as
low energies as those obtained by gradient flow.  The main advantage is
that
the fields and their currents  can be calculated by numerical integration
of ordinary differential equations, whereas the gradient flow  approach
requires
 the numerical solution of a partial differential equation. Also, there is
an
explicit separation  parameter $\rho$ in the instanton data, which is a
convenience.
A disadvantage of the instanton approach is that it requires  the pions
to be  regarded as massless. Other approaches can deal with pions which
have their physical mass. It would certainly be worthwhile to work with
$M_{10}$ as defined by the gradient flow, and compare with the results
here.

It is worth comparing   the picture of the two nucleon interaction that emerges
from the Skyrme
model with conventional nuclear potential models.
The basic difference is that no subset of the coordinates on $M_{10}$
can be identified simply with the positions of the two nucleons. While
the  two Skyrmions are separated one can identify two points where
the energy density (or baryon number density) is maximal, but when they
coalesce
these points disappear as the Skyrmions lose their identities. The energy
density of a  toroidal configuration is maximal on a circle.
Related to this is  the fact that,
in the attractive channel, two Skyrmions are never closer together
than in a toroidal configuration, and we shall define
below a separation parameter $\rho$ lying in the range $[\rho_0,\infty ]$,
where $\rho_0$ corresponds to the toroidal configurations, and is positive.
 When Skyrmions approach head-on they
pass through a toroidal configuration, scatter through $90^{\circ}$ and
separate
again.
  The radial part of the deuteron wavefunction $u(\rho)$
satisfies a radial Schr\"odinger equation on the interval
$[\rho_0,\infty)$.
Since $M_{10}$ is smooth at $\rho_0$, the boundary condition there
 is that $u$ is  finite and $du/d \rho =0$.
 In fact, $u(\rho)$ is maximal at $\rho_0$, which is where
not only the potential   $V$, but even the effective potential
$V_{\mbox{\tiny eff}}$ occurring in the radial equation is deepest.
This is quite different from a point-particle description of nucleons where
a hard core potential is required  to keep the nucleons from being close
together. With a hard core, the  boundary condition is $u(0)=0$, or
if the core potential is infinite for $\rho < \rho_1$, then $u(\rho_1)=0$.
In this situation, the wavefunction is usually maximal outside  the point
where
the potential is deepest. As a consequence, a stronger attraction is needed
to compensate. The Skyrme model seems to solve  the usual problem of a
`medium
range central attraction'  through the geometry of $M_{10}$.

As an aside,
we  point out that
the quantisation of a ten-dimensional family of two-Skyrmions was also
considered
in \cite{VWWW}.
The ten-dimensional family, like our manifold $M_{10}$, was obtained by
acting with translations, rotations and iso-rotations on a one-parameter
family of fields which interpolates between the toroidal configuration
and two-Skyrmions which are approximately of the product form, but have
additional reflection symmetries. These reflection symmetries also play
a role in our analysis, but our quantisation  scheme differs from that
of \cite{VWWW}, which
 appears incorrect. In \cite{VWWW}
an effective radial potential between Skyrmions was calculated  on the
interval $[0,\infty)$. The wavefunction was required to vanish at $\rho
=0$,
and there was (by a long way) no bound state.  We find a bound state with
a very similar potential because of our different (and we believe correct)
boundary conditions.

There is well-defined angular momentum in our formalism, but because
the Skyrmions are extended objects and interact, there is  no well-defined
split into orbital and spin parts of the angular momentum. Consequently
 we cannot perform the usual decomposition of the deuteron wavefunction
into
s-wave and d-wave parts. Nevertheless we can calculate physical quantities
related to this splitting, like the electric quadrupole moment.

The paper is organised as follows. Section 2 contains  a review of the
Skyrme model
 and its  symmetries. In section 3 we review how Skyrme fields are obtained
from instantons, and in particular how 2-Skyrmion attractive channel fields
are obtained and parametrised.
Section 4 describes the topological structure and the symmetries of
the manifold $M_{10}$.
The form of the Skyrme Lagrangian restricted
to attractive channel fields is described in section 5. The Lagrangian
depends on  nine functions of the Skyrmion separation parameter $\rho$.
These are calculated numerically, but for large values of $\rho$ we also
have analytic formulae, which provide a useful check.
 The appendix contains details of the numerical
methods used.
In section 6, the quantum  Hamiltonian in the attractive channel is derived
from the Lagrangian, and it is explained, using angular momentum analysis,
how the stationary Schr\"odinger equation reduces to a radial equation.
The bound states of this radial equation are discussed in section 7, and
the lowest physically allowed state is identified with the deuteron.
In section 8, various physical properties of our deuteron state
are calculated and compared with the values obtained by Braaten and Carson,
and with experiment. Some of the results are satisfactory, others less so.
For the first time a sensible binding energy of order  5 MeV  is obtained
in
 a Skyrme model calculation. The deuteron has  the right size, too, but
its electric quadrupole moment is too large and its magnetic dipole moment
is
too small.  We conclude, in section 9, with some remarks on how one might
correct our calculations on $M_{10}$ without going to a full quantisation
on the manifold $M_{12}$. Ideally, however, quantum  mechanics on $M_{12}$
is what should be done next.

\section{The Skyrme Model}

We set the speed of light to 1 and use the
metric  diag(1,-1,-,1,-1) on Minkowski space.
Points in Minkowski space are written as $(t,\bx)$ with  coordinates
$x^{\mu}$,
$\mu =0,1,2,3$, and the Einstein summation convention is used throughout.
The basic field  of the Skyrme model
is the $SU(2)$-valued field $U(t,\bx)$,  which
can also be expressed in terms of the pion  fields:
\bea
\label{pion}
U(t,\bx) = \sigma(t,\bx) + i\bpi(t,\bx)\cd\btau
\eea
where $\tau_1,\tau_2,\tau_3$ are the Pauli matrices and $\sigma^2 + \bpi^2
=1$.
The equations of motion for $U$ are the Euler-Lagrange equations
 derived from the Lagrangian density
\bea
\label{lag}
{\cal L} = -{F_{\pi}^2 \over 16} \tr( R_{\mu} R^{\mu}) +
{1\over 32 e^2} \tr( [R_{\mu},R_{\nu}]
[R^{\mu},R^{\nu}]).
\eea
The right-currents $R_{\mu}$ are defined via
\bea
R_{\mu} = (\partial_{\mu} U) U^{\d},
\eea
where $\partial_{\mu}$ denotes partial differentiation with respect to
$x_{\mu}$.
The constants $F_{\pi}$ and $e$ are free parameters of the Skyrme model.
We found it convenient to remove them from our calculations by
using  $F_{\pi}/4e$ as the  unit of energy and
 $2/e F_{\pi}$ as the unit of length.
Thus we work in geometrical units, and the Lagrangian density takes the
form
\bea
{\cal L} = -{1\over 2} \tr( R_{\mu} R^{\mu}) + {1\over 16} \tr(
[R_{\mu},R_{\nu}]
[R^{\mu},R^{\nu}]).
\eea

The parameters $F_{\pi}$ and $e$ can be fixed in a number of ways.
Most authors adopt  the approach of \cite{ANW} and  \cite{AN}
  where  $F_{\pi}$ and $e$ are tuned
to  reproduce the masses of the proton and the  delta resonance
without  and with the physical pion mass respectively.
All  papers written so far   on the deuteron in the Skyrme model
take into account the physical pion mass and use the  set of parameters
 given in \cite{AN}. For ease of comparison we will
 follow that practice, although
the pion mass
is assumed to be zero here.
Thus, with the values of $F_{\pi}$ and $e$ as in \cite{AN}, our units
are related to conventional units via
\bea
\label{units}
{F_{\pi} \over 4e} = 5.58\,\mbox{MeV} \qquad \mbox{and}
 \qquad {2\over eF_{\pi}}=0.755\,\mbox{fm}.
\eea
Since
\bea
\label{hbar}
\hbar = 197.3 \mbox{MeV fm}= 46.8 \left({F_{\pi} \over 4e}\right)
\left( {2\over eF_{\pi}}\right),
\eea
it  follows that, in our units where $  F_{\pi} / 4e=  2 / eF_{\pi}=1$,
$\hbar  =46.8$.

It is useful to think of the Skyrme model
as an infinite-dimensional Lagrangian system
whose  configuration
space $Q$ is the
space of
maps
\bea
U : {\bf R}^3 \mapsto    SU(2)
\eea
which obey
\bea
\label{inf}
\lim_{|\bx|\rightarrow \infty} U(\bx) =\bone .
\eea
The Lagrangian $L = \int d^3 x {\cal L}$ has the usual form
$L= T-V$, where
the $T$ is the kinetic  energy
\bea
\label{KE}
T = \int d^3 x\left\{ -{1\over 2} \tr (R_0 R_0)  - {1\over 8}
\tr ([R_i,R_0][R_i,R_0])\right\}
\eea
and
$V$ is the potential energy
\bea
\label{PE}
V= \int d^3 x\left\{ -{1\over 2} \tr (R_i R_i)  - {1\over 16}
\tr ([R_i,R_j][R_i,R_j])\right\}.
\eea

The condition (\ref{inf}) is imposed in order to ensure that   the elements
of $Q$ have finite energy.
It is important topologically because it  allows one to compactify
${\bf R}^3$ to $S^3$ and to
regard an element  $U$ of $Q$ as a map $S^3 \mapsto SU(2)\cong S^3$.
Thus $\pi_0(Q)=\pi_3(S^3) ={\bf Z}$, showing that
$Q$ is not connected  but has components $Q_B$ labelled by an integer
$B$ which  is the degree of any element of  $Q_B$.
Physically, the integer $B$ is interpreted as the baryon number of  a
Skyrme
field. It can be calculated by integrating the zeroth component of
the (conserved)
baryon number current
\bea
\label{bcurrent}
B^{\mu} =  { \epsilon^{\mu\nu\alpha\beta}\over 24
\pi^2}\tr(R_{\nu}R_{\alpha}
R_{\beta}),
\eea
where we use the convention $\epsilon_{0123}= -\epsilon^{0123} =1$.
Thus,
\bea
\label{bdensity}
 B^0 = -B_0 = - {\epsilon_{ijk}\over 24 \pi^2} \tr(R_iR_jR_k)
=-{1\over 8\pi^2} \tr([R_1,R_2]R_3)
\eea
and
\bea B^i = B_i ={\epsilon_{ijk} \over 8 \pi^2}\tr(R_jR_kR_0).
\eea
Then
the baryon number is
\bea
\label{bnumber}
B = \int d^3 x B^0.
\eea

The fundamental group of $Q$ is also important here. In general, the
fundamental
group of a space depends on the choice of a base point, but it is explained
in \cite{Nico} that
\bea
\pi_1(Q)= \pi_1(Q_B) ={\bf Z}_2
\eea
for any $B \in{\bf Z}$.  The group $\pi_1(Q)$ is
generated by the rotation of a single Skyrmion by $2\pi$. More generally,
it
is shown in \cite{Nico} that the loop generated by rotating an element
of $Q_B$ by $2\pi$ is contractible if $B$ is even and not contractible if
$B$ is odd. A further  important topological result, due to
Finkelstein and Rubinstein \cite{FR}, is that   the rotation of
one
Skyrmion by $2\pi$ and  the exchange of two Skyrmions are homotopic paths in
$Q_B$, for $|B|\geq 2$.

We will use the following
conventions  when describing Skyrme fields.
We  refer to the domain $ {\bf R}^3$ of  a static Skyrme field
as   physical space
and to the range $SU(2)$ as iso-space.
We write $\{\be_1,\be_2,\be_3\}$ for   the  canonical orthonormal basis
 of   ${\bf R}^3$   ($\be_1=(1,0,0)$ etc.)  and in the decomposition
  (\ref{pion}) of elements of $SU(2)$  we refer to $i\tau_a$
   ($a=1,2,3$) as the $a$-th axis in  iso-space.
$G$ stands for  an   $SU(2)$ matrix and  $D(G)$ for
the $SO(3)$ matrix associated to  $G$ via
\bea
D(G)_{ab}= \frac12 \tr (\tau_a G \tau_b G^{\d}).
\eea

For fixed $B$, the symmetry group of the Lagrangian system with
configuration space
$Q_B$ and Lagrangian $L$ is
\bea
\label{symgroup}
 P\times SO(3)^I \times E_3.
\eea
Here $P$ is the combined parity operation in space and iso-space,
\bea
\label{inversion}
P: U(\bx) \mapsto U^{\d}(-\bx),
\eea
and $SO(3)^I$ is the group of iso-rotations. Its action can be
written in terms of an $SU(2)$ matrix $C$ as
\bea
U(\bx) \mapsto CU(\bx)C^{\d}
\eea
or, in terms of the pion fields,
\bea
\bpi(\bx) \mapsto D(C)\bpi(\bx).
\eea
Finally, the euclidean group  $E_3$ is the semidirect product of the
spatial
rotation group $SO(3)^J$ and the group ${\bf R}^3$ of translations.
An element $(D(G),\bS)\in E_3$ acts  on
a vector $\bx$  according to
\bea
\bx \mapsto D(G) \bx + \bS
\eea
and  on Skyrme fields via pull-back:
\bea
U(\bx) \mapsto U(D(G)^{-1} (\bx -\bS)).
\eea

The  group (\ref{symgroup}) has discrete subgroups which will be important
for us.
We therefore
introduce the notation $O_{ai}$ ($a,i\in{1,2,3}$)  for simultaneous
 rotations by $\pi$
around the $a$-th axis in iso-space and the $i$-th axis in physical space.
It will also be useful to write $O_{0i}$ for the  rotation by $\pi$
about
the $i$-th axis in physical space only and $O_{a0}$ for the
rotation by $\pi$
about the $a$-th axis in  iso-space only. Finally we  define
hyperplane reflections in space and iso-space
$P_{ai}= PO_{ai}$.

For the discussion of the kinetic energy we need carefully to define
angular
 velocities:
the   left-invariant or body-fixed angular velocity for spatial rotations
$\bom$  is defined via
\bea
\bom \cd \bt = G^{\d} \dot G,
\eea
where $t_i = -{i\over 2} \tau_i$ (so that $[t_i,t_j]=\varepsilon_{ijk}t_k$)
and  the dot denotes differentiation with respect to time.
Equivalently, in terms of $D(G)$:
\bea
\bom \cd \bs  = D(G)^{-1} \dot D(G),
\eea
where the $s_i$ are $3 \times 3$ matrices with components $(s_i)_{jk} =
-\varepsilon_{ijk}$. They also satisfy $[s_i,s_j] =\varepsilon_{ijk}s_k$.
The corresponding right-invariant or space-fixed angular velocity is less
important for us; it is given by $D(G) \bom$.

Similarly one defines the body-fixed angular velocity
$\bOm$ for iso-rotations
via
\bea
\bOm \cd \bt = C^{\d} \dot C.
\eea
The  corresponding space-fixed angular velocity is again given by
$D(C) \bOm$.
The angular momenta associated to the angular velocities in space and
iso-space will be discussed in section 6.

\section{Attractive Channel Skyrme Fields from Instantons}

The idea that Skyrme fields can be generated from Yang-Mills instantons
was proposed in \cite{AM1} and subsequently developed in a number of
papers, including
\cite{AM2} and \cite{AM3}. The attractive channel fields generated from
instantons were
also studied in \cite{HGOA}.

Briefly, consider the fourth component  $A_4$ of a self-dual $SU(2)$
Yang-Mills
field, or instanton, on ${\bf R}^4$, and define the  holonomy along all
lines parallel to the $x_4$-axis
\bea
\label{hol}
U(\bx) = {\cal P} \mbox{exp} - \int_{-\infty}^{\infty} A_4(\bx,x_4)dx_4
\eea
(the right hand side is a path-ordered  exponential). The formula
(\ref{hol})
defines an $SU(2)$-valued field which can be regarded as a Skyrme field on
${\bf R}^3$. Provided $A_4$ decays sufficiently rapidly as $|\bx|
\rightarrow
\infty$, $U(\bx)$ satisfies the asymptotic condition (\ref{inf}).
Moreover, the degree, or baryon number of $U$ equals the instanton
number or charge \cite{AM3}.

The expression (\ref{hol}) is formal. It is computed by solving the
differential
equation
\bea
\label{holeq}
\partial_4 \tilde{U} = -A_4 \tilde{U},
\eea
where $\tilde{U}$ is defined on ${\bf R}^4$. One imposes the boundary
condition
\bea
\label{minf}
\lim_{x_4 \rightarrow -\infty} \tilde{U}(\bx, x_4) = \bone
\eea
and defines  the Skyrme field $U(\bx)$ via
\bea
\label{pinf}
U(\bx) = \lim_{x_4 \rightarrow \infty} \tilde{U}(\bx,x_4).
\eea
For  details of the numerical implementation of these steps we  refer
the
reader to the appendix, and also \cite{LM}.

The simplest Skyrme field one obtains this way (apart from the
vacuum) is the hedgehog Skyrme field of degree 1. This is generated by a
unit
charge instanton centred at the origin. As shown in \cite{AM1} the equation
(\ref{holeq})  can be solved analytically in this case, and gives the
Skyrme field
\bea
\label{hedge}
U_H(\bx)=\exp i f(r)\hat\bx\cd\btau
\eea
where
\bea
\label{hedgehog}
f(r) = \pi \left( 1- (1+ {\lambda \over r^2})^{-{1\over 2}}\right).
\eea
The scale parameter $\lambda$ is chosen to minimise the energy of the
field,
which is interpreted as the Skyrmion mass. The minimum is obtained for
$\lambda =
2.11$; Then the mass is
$M= 1.243\cd 12 \pi^2 = 147.2$
and the  moment of inertia is $\Lambda = 8.363\cd 16\pi/3=140.1$
 (normalised so that
the kinetic energy of a hedgehog spinning  with frequency $\omega$ is
${1\over 2}
\Lambda \omega^2$). Using the formula $M + 3\hbar^2/8\Lambda$ \cite{ANW},
it  follows that the prediction for the nucleon mass is
$M=153.1$, which is 854.1 MeV in physical units.
This rather small value results from the combination of the instanton
generated profile function (\ref{hedgehog}) with the choice of units
(\ref{units}) designed to reproduce the  experimental values for the
 nucleon and delta masses
when the physical pion mass is  taken into account. However, the absolute
value of the nucleon mass (and later the deuteron mass) is not important
here. We are mainly interested in the deuteron's binding energy, which
is the difference  between its own mass and
the  mass of two free nucleons.
For later use we also note that  $f$ has the simple form
\bea
f(r) \sim {p\over r^2}
\eea
when $r$ is large.
The constant $p$ (which may be interpreted as the Skyrmion's pion dipole
strength) is $ \pi\lambda/2$.

The attractive channel two-Skyrmion fields, generated  by instantons, were
discussed in detail in \cite{AM3}. There is also an analytic formula
for $A_4$, given by Jackiw, Nohl and Rebbi (JNR) in \cite{JNR},
but the equation (\ref{holeq}), which gives the Skyrme fields,
must still be solved numerically. The JNR formula is somewhat gauge
dependent
and ungeometrical, so before giving it we recall Hartshorne's geometrical
description of charge two instantons. Hartshorne showed that associated
to any charge two $SU(2)$ instanton in ${\bf R}^4$ there is a circle and an
ellipse  in ${\bf R}^4$ (the circle can degenerate to a straight line but
this will
not be relevant here). The ellipse is in the same plane as the circle, and
interior to it. Moreover, the ellipse satisfies the Poncelet condition;
 that is, there exists a triangle with vertices on the circle whose sides
are tangent to the ellipse. Poncelet's theorem states that if one such
triangle exists, then there are infinitely many and any point on the
circle can be taken as one vertex.

Two well-separated Skyrmions maximally attract if one is rotated relative
to the other by $\pi$ about a line perpendicular  to the line
joining them.
 The attractive channel interpolates  between such Skyrmion pairs
and the toroidal two-Skyrmion, where the Skyrmions coalesce
into the  minimal energy configuration. The attractive channel is modelled
by a subset of the Skyrme fields generated by charge two instantons.
This is the subset where the Hartshorne circle lies in a spatial plane in
${\bf R}^4$ (orthogonal to the $x_4$-axis) and where the ellipse is concentric
with the circle. Without loss of  generality, we may suppose the common
centre
is at the origin in $\bf R^4$, that the circle is in the $x_1x_2$-plane,
and
that the ellipse has its major axis along the 1-axis, see figure  1.
In this standard form
there are two  parameters in the Hartshorne data for attractive
channel fields. These may be taken to be the radius $L$ of the circle
and the length $L_1$  of the semi-major axis of the ellipse. We denote the
length
of the semi-minor axis by $L_2$.  The Poncelet condition is that $L_1+L_2=L$.
The configuration in figure 1 is  rather symmetric, and this manifests
itself
in the symmetries of the corresponding Skyrme fields, which we
will discuss in section 4.

When $L_1 \gg L_2$, the instanton associated with the circle and ellipse of
figure 1
generates a Skyrme field consisting of two well-separated Skyrmions,
centred
near the ends of the ellipse (the $x_4$-coordinate is irrelevant now).
 When $L_1=L_2={1\over 2}L$, the ellipse becomes a circle, so there is
  higher symmetry, and the  instanton generates a toroidal Skyrme field.

To progress, we need the explicit JNR formula for $A_4$ for the instantons
whose Hartshorne data is as in figure 1. To this  end we choose one
triangle
of the Poncelet porism; following Hosaka {\it et  al. } \cite{HGOA} we choose
the
isosceles triangle shown.  The JNR data consists of the points $X_1,
X_2,X_3$
in ${\bf R}^4$ as shown and associated weights
$\lambda_1,\lambda_2,\lambda_3$.
These can be rescaled so that $\lambda_2=\lambda_3=1$. It is natural
to define the angle $\vartheta =
\arcsin (1/( 1+\lambda_1))$, whose geometrical significance is shown in
figure 1. In JNR terms, the essential parameters of the instantons are the
radius $L$ of the circle and the angle  $\vartheta$, which lies in the range
$(0,{\pi \over 6}]$. The semi-major axis of the ellipse is then $L_1=(1-\sin
\vartheta)L$.
The JNR gauge potential in ${\bf R} ^4$ has  a
 fourth component of the form
\bea
\label{A4explicit}
A_4=\frac i2\,\partial_i \ln \nu(x) \tau_i,
\eea
where $\nu$ is the following  function of $x=(\bx,x_4)\in {\bf R}^4$:
\bea
\label{Rhoexplicit}
\nu(x)=\sum_{n=1}^3\,{\lambda_n\over |x-X_n|^2}\,.
\eea
There is a slight technical complication in that the JNR gauge
 potentials do not
decay sufficiently rapidly at large distances  to
extend smoothly from ${\bf R}^4$ to $S^4$; it turns out that an extra
factor
of $-1$ must be  attached to $U$ in (\ref{pinf}), see also \cite{AM3}.
The Skyrme fields in the attractive channel are then given by (\ref{holeq})
- (\ref{pinf}).

The Hartshorne data in the standard form  described above have
two parameters, $L$ and $\vartheta$,  and acting with spatial translations,
rotations and  iso-rotations on  the Skyrme fields generated
from the corresponding instantons one obtains an eleven-dimensional family
of Skyrme fields, which is not desired. Instead of retaining $L$ and
$\vartheta$ as independent parameters we therefore fix $L$, for each value
of $\vartheta$, to the value which minimises the Skyrme potential energy.
This fixes the scale of the Skyrmions in the configuration, so that for
$\vartheta \approx 0$  we obtain the product of  two well-separated
hedgehog
Skyrmions with
the scale factor $\lambda =2.11$ as in  (\ref{hedgehog}) and oriented
so that the attraction is maximal. In the limit $\vartheta \rightarrow
0$, the energy of this configuration tends to $2M=294.4$,  twice the energy
of a
hedgehog Skyrmion.
For $\vartheta = \pi/6$ we  obtain a toroidal
Skyrme field whose energy is $1.19 \cd 24 \pi^2 =282.0$.

As a separation parameter we shall use $\rho = 2L (1-\sin\vartheta) = 2
L_1$. This
 makes  physical sense: $\rho$ is positive when $\vartheta = \pi/6$, taking
the
value $\rho_0 =1.71$, and it tends to infinity as $\vartheta \rightarrow
0$.
For small $\vartheta$, $\rho$ is an accurate estimate of the distance
between
 the Skyrmion centres (for example, $2L$ is not so good).
The field approaches a product of hedgehog fields with this separation,
as can be seen by considering another triangle in the Poncelet porism,
namely one which is isosceles and has one vertex on the 1-axis (see also
\cite{AM3}).
Mathematically, our definition of separation is a little awkward,
as $L$ is a
function of $\vartheta$ which is determined numerically, but this
causes few problems in practice.

To sum up, starting with the standard  Hartshorne data  depicted in
figure 1 we generate a 1-parameter family of  attractive channel
Skyrme fields in standard orientation, which we denote by $\hat
U(\rho,\bx)$.
The iso-orientation of $\hat U(\rho,\bx)$ is chosen so that for large
$\rho$ it is approximately a product of two well-separated hedgehog
Skyrmions
on the 1-axis with one hedgehog rotated relative to the other
by $\pi$ about the 3-axis:
\bea
\label{prodan}
\hat U(\rho,\bx) \approx U_H (\bx +\frac12 \rho \be_1)
\tau_3  U_H(\bx-\frac12 \rho   \be_1 )\tau_3.
\eea
When $\rho=\rho_0$, $\hat U(\rho,\bx)$ is the toroidal Skyrme field,
 centred
at the origin and   with the axis of
the torus along the $3$-axis.

\section{Topology and Symmetries of the Manifold of
Attractive \newline Channel Fields}

We define the manifold $M_{10}$ of attractive channel fields  as
the orbit of the family of  standard fields $\hat U(\rho,\bx)$ under
the symmetry group (\ref{symgroup}). For most of this paper we  will not
be concerned with
 overall translations of the standard field $\hat U(\rho,\bx)$.
Thus we also  define the manifold $M_7$ of centred
attractive channel fields   as the quotient of $M_{10}$ by the translation
group ${\bf R}^3$. Alternatively one can think of $M_7$ as the orbit
 of the family of  standard fields $\hat U(\rho,\bx)$ under
\bea
{\cal G} = SO(3)^I\times SO(3)^J\times P.
\eea
For fixed $\rho$, the orbit is diffeomorphic to the quotient of ${\cal G}$
by the isotropy group of its action on $\hat U(\rho,\bx)$.

The isotropy group consists, by definition, of all elements of ${\cal G}$
which leave $\hat U(\rho,\bx)$  invariant. The standard Hartshorne
data in figure 1 is clearly invariant under
 reflections in the $x_1x_2$, $x_2x_3$ and $x_1x_3$ planes, but the
Skyrme fields $\hat U(\rho,\bx)$
are only invariant under certain combinations of these reflections  with
reflections in iso-space.
Explicitly these combinations are:
\bea
\label{dissym}
P_{21}: (\pi_1,\pi_2,\pi_3) \mapsto (\pi_1,-\pi_2,\pi_3) \quad \mbox{and}
\quad
(x_1,x_2,x_3) \mapsto (-x_1,x_2,x_3) \nonumber \\
P_{22}: (\pi_1,\pi_2,\pi_3) \mapsto (\pi_1,-\pi_2,\pi_3) \quad \mbox{and}
\quad
(x_1,x_2,x_3) \mapsto (x_1,-x_2,x_3) \nonumber \\
P_{33}: (\pi_1,\pi_2,\pi_3) \mapsto (\pi_1,\pi_2,-\pi_3) \quad \mbox{and}
\quad
(x_1,x_2,x_3) \mapsto (x_1,x_2,-x_3) .
\eea
The group generated by  these maps is an abelian subgroup of
(\ref{symgroup})
of order 8 which we
denote by ${\cal G}_8$. Its elements are
\bea
{\cal G}_8 = \{ 1,O_{11},O_{12},O_{03},PO_{30},P_{21},P_{22},P_{33} \}.
\eea
It is easy to check that
\bea
\mbox{Vier} =\{ 1,O_{11},O_{12},O_{03} \}
\eea
is a  subgroup which is isomorphic to the viergruppe.  Using
\bea
PO_{30}P_{21}= O_{11} \nonumber \\
PO_{30}P_{22}= O_{12} \nonumber \\
PO_{30}P_{33}= O_{03}
\eea
one shows further that there is an  isomorphism
\bea
{\cal G}_8 \cong \mbox{Vier} \times  \mbox{\bf Z}_2,
\eea
where {\bf Z}$_2$ =$\{1,-1\}$,
which identifies $PO_{30}\in {\cal G}_8 $ with $(1,-1)\in \mbox{Vier}
\times  \mbox{\bf Z}_2$.
It follows that
for $\rho >\rho_0$  the isotropy group of $\hat U(\rho,\bx)$ is
${\cal G}_8$ and  that the orbit  under $\cal G$ is diffeomorphic to
\bea
\label{orbit}
\left(SO(3)^I\times SO(3)^J \right)/ \mbox{Vier}.
\eea

When $\rho=\rho_0$, there is an additional invariance.
The field $\hat U(\rho_0,\bx)$ is invariant under spatial rotations about
the 3-axis
by  some angle $\chi$ and simultaneous iso-rotation about the 3-axis by
$2\chi$.
Explicitly
\bea
\label{so2}
e^{-i\chi \tau_3}\hat U(\rho_0,D(e^{{i \chi \over 2}\tau_3})\bx)
e^{i\chi\tau_3} =\hat U(\rho_0,\bx),\qquad \chi\in[0,2\pi).
\eea
 Such transformations form an $SO(2)$
subgroup  of $SO(3)^I \times SO(3)^J$ which we denote by ${\cal H}$.
It follows that the isotropy group of $\hat U(\rho_0,\bx)$ is  the
semi-direct
 product of    $\cal H$ with the group consisting of
  $\{1,P_{22},P_{33},O_{11}\}$.
Thus
the orbit under $\cal G$ is diffeomorphic to
\bea
\left(SO(3)^I\times SO(3)^J \right) / O(2).
\eea
Here $O(2)$ is the semi-direct product of $\cal H$ with the group
$\{1,O_{11}\}$. Note that the orbit under the action of ${\cal G}$ is
six-dimensional
if $\rho > \rho_0$ but only five-dimensional if $\rho=\rho_0$.

Concretely,  any  field  in $M_{10}$
 can be written in terms
of $SU(2)$ matrices
$C$ and $G$ and a translation vector $\bS$ as
\bea
\label{ansatz}
 C \hat U(\rho, D(G)^{-1}(\bx -\bS)) C^{\d}.
\eea
For $\rho >\rho_0$, $D(C),D(G)$ and $\bS$ can be used to parametrise
the fields, but
we should identify points related by the right action of $O_{03}$
and $O_{11}$:
\bea
\label{identify}
(D(C),D(G))
& \sim& (D(C),D(G) O_{03})
\nonumber \\
(D(C),D(G))
&\sim& (D(C) O_{10},D(G) O_{01}).
\eea
It follows, as a consequence of the group structure,  that points related
by
the right action of $O_{12}$ are also identified.
For $\rho =\rho_0$ we identify points related by the right action $O_{11}$
and
by the right action  of elements in $\cal H$:
\bea
\label{so2id}
(D(C),D(G))
&\sim &(D(C) O_{10},D(G) O_{01})\nonumber\\
(D(C),D(G)) &\sim& (D(Ce^{-i\chi \tau_3}),D(Ge^{-{i \chi \over 2}\tau_3})).
\eea

The $\cal G$-orbit structure of  $M_7$
can be used to coordinatise $M_7$ in terms of the separation
parameter $\rho$ and Euler angles $(\phi,\theta, \psi)$ for $SO(3)^J$
and  $(\alpha, \beta,\gamma)$ for $SO(3)^I$.  More precisely we write
an element $D(G)$  of  $SO(3)^J$ as
\bea
\label{Euler}
D(G) =e^{\phi s_3} e^{\theta s_2} e^{\psi s_3}
\eea
where $s_i$ are the generators of $SO(3)$ defined in section 2.
Similarly an element $D(C)$ of $SO(3)^I$ can be written as
\bea
D(C) = e^{\alpha s_3}e^{\beta s_2}e^{\gamma s_3}.
\eea
The range of  the Euler angles is $\phi,\psi,\alpha,\gamma \in[0,2\pi)$ and
$\theta, \beta \in [0,\pi)$. However, points related by  the maps
$O_{03},O_{11},
O_{12}$  should be identified. These maps now take the form
\bea
\label{I13TN}
\begin{array}{llll}
O_{03}:& \beta \mapsto \beta,&
         \alpha \mapsto \alpha,&
       \gamma \mapsto \gamma,\\
  &      \theta \mapsto \theta, &
        \phi \mapsto  \phi , &
        \psi \mapsto  \psi+\pi \\
O_{11}: & \beta \mapsto \pi -\beta,&
        \alpha \mapsto \alpha+\pi,&
       \gamma \mapsto -\gamma ,\\
     & \theta \mapsto \pi-\theta, &
        \phi \mapsto  \phi +\pi, &
        \psi \mapsto  -\psi \\
O_{12} :& \beta \mapsto \pi-\beta, &
         \alpha \mapsto \alpha +\pi,&
          \gamma \mapsto   -\gamma, \\
     &    \theta \mapsto  \pi -\theta, &
        \phi \mapsto  \phi+\pi ,    &
        \psi \mapsto \pi -\psi.
\end{array}
\eea

Using the orbit structure of  $M_7$ it is not difficult  to understand
its   homotopy
structure.
This will be important in the quantum theory. All the results that are
required
 here were already
derived in \cite{VWWW}, but it is useful to briefly state and prove them
using our notation.
Since $M_7$ is connected it is sufficient to
consider paths which begin and end at the toroidal configuration
$\hat U(\rho_0,\bx)$.

Our first claim is  that loops generated by either
spatial or iso-spatial rotations of $\hat U(\rho,\bx)$ by $2 \pi$
are
contractible in $M_7$.
To see this we first note that rotations by $2\pi$ about any two
axes are homotopic to each other; thus we can specify a convenient
axis without loss  of generality.
The claim is clearly true for a loop generated by a spatial rotation
of $\hat U(\rho_0,\bx)$ by $2 \pi$ about  the 3-axis, since
such a rotation  is equivalent to an iso-rotation by
$4 \pi$  about the 3-axis. Since rotations by $4 \pi$
are homotopic to
the identity  the loop is contractible.
To show the  contractibility of the loop generated by an iso-rotation by
$2 \pi$ we also initially consider  a rotation about the
3-axis. Then  we smoothly increase $\rho$
until    $\hat U(\rho,\bx)$ is approximately the product of two
 well-separated
 hedgehog fields as in (\ref{prodan}).
Since the hedgehogs are rotated relative to each  other about the
3-axis,  iso-rotating the whole configuration  by $2\pi$
 about the 3-axis is the same
as iso-rotating each of the  hedgehogs by the same amount about the 3-axis.
However, by our earlier remark  this loop is homotopic to rotating
each of hedgehogs by $2\pi$ about the 1-axis.
Moreover, for hedgehogs   iso-spatial rotations are the same as spatial
rotations.
 Then,  since the two hedgehogs are  separated along
 the 1-axis, a
spatial rotation of each by $2\pi$ about the 1-axis
  is  equivalent to
   an overall spatial rotation by $2\pi$  about the 1-axis.
   The contractibility of the loop then
  follows from  our earlier result that
  a spatial rotation
by $2 \pi$ is contractible.

Our second claim is that all loops in $M_7$ are either contractible or
homotopic to  a rotation of one of the  hedgehog fields in the  product
ansatz
(\ref{prodan}). It then follows that $\pi_1(M_7)={\bf Z}_2$, so
$M_7$ has the same  fundamental group as  the space $Q_2$ of all Skyrme
fields
of degree 2.
To prove the  claim we have to consider the paths which connect
points identified by $O_{03}$, $O_{11}$ and $O_{12}$.
First consider the path $\Gamma_{03}$
 in $M_7$ obtained by  performing a spatial
rotation  of $\hat U(\rho_0,\bx)$ by $\pi$ about  the 3-axis. The
map
$O_{03}$ identifies the endpoint with the starting point, so the path
is closed. However, for toroidal configurations a  spatial
rotation by $\pi$ about the 3-axis is equivalent to an
iso-rotation by $2 \pi$  about the 3-axis. Hence the path is
contractible
by our first claim.
The second path we consider is obtained by rotating $\hat U(\rho_0,\bx)$
in both space and iso-space by $\pi$ about the 1-axes.
 This path, denoted $\Gamma_{11}$, is closed because
 the map $O_{11}$ identifies the endpoint
with the starting point. By increasing $\rho$  so that $\hat U(\rho,\bx)$  is
of the
product form (\ref{prodan}) this loop can be deformed into the following
\bea
 e^{-{i\over 2}\chi\tau_1}U_H (D( e^{{i\over 2}\chi\tau_1})\bx +
 \frac12 \rho \be_1))
\tau_3  U_H(D( e^{{i\over 2}\chi\tau_1})\bx-\frac12 \rho  \be_1 )\tau_3
 e^{{i\over 2}\chi\tau_1}
\nonumber \\
\qquad \qquad \qquad =
U_H (\bx +\frac12 \rho \be_1))
\tau_3  U_H(D( e^{i\chi\tau_1})\bx-\frac12 \rho   \be_1 )\tau_3 ,
 \qquad \qquad  \chi \in[0,\pi).
\eea
Hence it is homotopic to a rotation by $2 \pi$  of one of the two
Skyrmions in the product ansatz. We know that such a loop is not
contractible in $Q_2$ and,
{\it a fortiori}, it is not contractible in $M_7$. It follows  also that the
loop $\Gamma_{11}^2$ is homotopic to a rotation by $4 \pi$ of one  of
the Skyrmions
and hence contractible.
Finally there is the loop $\Gamma_{12}$  associated with the identification
$O_{12}$. It is obtained by performing a spatial rotation by $\pi$
about the
2-axis and an iso-rotation by $\pi$ about the 1-axis.  It
follows from  $O_{12}=
O_{11}O_{03}$ that $\Gamma_{12}=\Gamma_{11}\Gamma_{03}$. Hence
$\Gamma_{12}$
is in the same homotopy class as $\Gamma_{11}$ and not contractible.

To summarise, loops in $M_7$ are either contractible or homotopic to a
rotation
of one Skyrmion by $2 \pi$. Our discussion of the loops
$\Gamma_{11},\Gamma_{03}$
and $\Gamma_{12}$ suggests the following physical interpretation of the
identifications
 $O_{11},O_{03}$ and $O_{12}$ in  the asymptotic region of $M_7$
which describes well-separated Skyrmions: $O_{11}$ identifies
configurations
related by rotating one of the Skyrmions by $2 \pi$, $O_{12}$
identifies
configurations related by the exchange of the two Skyrmions and $O_{03}$
identifies configurations related by the exchange of the two Skyrmions and
the simultaneous rotation of each by $\pi$.
As mentioned in section 2,   the rotation of
one Skyrmion and  the exchange of two Skyrmions are homotopic paths in
$Q_2$.
We have seen that the two operations are also homotopic in $M_7$.
Note, however, that  the homotopy does  not  require  the creation
of a Skyrmion - anti-Skyrmion pair, as is  often claimed.
In fact, we have checked that  the baryon density for all fields
in $M_7$ is non-negative everywhere.

\section{The Lagrangian in the Attractive Channel}

The Lagrangian in the attractive channel is, by definition,
the restriction of the  Skyrme Lagrangian $L$
to $M_{10}$. We call  it $L_{\mbox{\tiny att}}$.
The  evaluation of potential energy $V$
gives a function  of $\rho$ only which we also denote by $V$.
To calculate the kinetic energy, denoted $T_{\mbox{\tiny att}}$,  we allow
  $C,G, \bS$ and $\rho$ in (\ref{ansatz})
  to vary with time.
We write
\bea
\label{cur}
\hat R_i = (\partial_i \hat U) \hat U^{\d}, \qquad
\hat R_{\rho} =(\partial_{\rho}\hat  U ) \hat  U^{\d}
\quad \mbox{and} \quad \by = D(G)^{-1}(\bx -\bS),
\eea
where $\partial_{\rho}$ denotes differentiation with respect to $\rho$;
we also introduce the notation  $\bRh$ for the vector with components $\Rh_i$.
We then find,
in terms of the angular velocities $\bom$ and $\bOm$ defined in section 2,
\bea
R_0 = C \left( [-\frac i2 \bOm\cd\btau, \hat U] \hat U^{\d} -
\bom \cd \by \times \bRh_y - \dot\bS\cd D(G) \bRh_y  + \dot \rho
\Rh_{\rho} \right) C^{\d}.
\eea
Here, all functions are evaluated at $\by$ and we   have written $\bRh_y$
to indicate that the differentiation should be carried out with respect
to $\by$.
Inserting this formula into (\ref{KE}), changing integration
variables to $\by$ ($d^3x =d^3 y$) and then changing the
name of the integration
variable from $\by$ back to $\bx$,  we  find that the kinetic
energy
can be calculated solely from the standard field $\hat U(\rho,\bx)$ and
its currents
$\hat R_i$ and $\hat R_{\rho}$.

The calculation  is  simplified
by  symmetries. It follows from the invariance of the  kinetic energy
$T_{\mbox{\tiny att}}$ under the  left  action of $\cal G$  that  it
 can be expressed as a (positive, symmetric) bilinear form  in $\dot \rho$
and the left-invariant angular velocities $\omega_i$ and $\Omega_a$
($i,a=1,2,3$).
The bilinear form is further restricted by the identifications
  (\ref{identify}) which imply that we should identify the
the left-invariant angular velocities $\bom$ and $\bOm$
 calculated from $D(C)$ and $D(G)$ with those calculated from the image
of $D(C)$ and $D(G)$
 under the right action of $O_{11},O_{12}$ and $O_{03}$. Explicitly,
 the angular velocities transform as follows:
\bea
\begin{array}{lll}
O_{03}&: (\Omega_1,\Omega_2,\Omega_3) \mapsto (\Omega_1,\Omega_2,\Omega_3)
\quad &\mbox{and} \quad (\omega_1,\omega_2,\omega_3) \mapsto
(-\omega_1,-\omega_2,\omega_3)  \nonumber \\
O_{11}&: (\Omega_1,\Omega_2,\Omega_3) \mapsto
(\Omega_1,-\Omega_2,-\Omega_3)
\quad &\mbox{and} \quad (\omega_1,\omega_2,\omega_3) \mapsto
(\omega_1,-\omega_2,-\omega_3) \nonumber \\
O_{12}&: (\Omega_1,\Omega_2,\Omega_3) \mapsto
(\Omega_1,-\Omega_2,-\Omega_3)
\quad &\mbox{and} \quad (\omega_1,\omega_2,\omega_3) \mapsto
(-\omega_1,\omega_2,-\omega_3).
\end{array}
\eea
The kinetic energy must respect these identifications,  which implies that
it is of the following form in the centre of mass frame ($\dot \bS=
\b0$):
\bea
\label{form}
T_{\mbox{\tiny att}} = {1\over2}\left( f^2 {\dot \rho}^2 + a^2 \omega_1^2 +
b^2 \omega_2^2
+ d^2 \omega_3^2  + A^2\Omega_1^2
+B^2 \Omega_2^2 + C^2\Omega_3^2 +2e \omega_3\Omega_3\right),
\eea
where $a,b,d,e,f,A,B$ and $C$ are functions of $\rho$ only.
Explicitly we find
\bea
\label{ff}
f^2 = \int d^3 x \left\{
- \tr(\Rh_{\rho} \Rh_{\rho}) -{1\over 4}
\tr([\Rh_{\rho},\Rh_i][\Rh_{\rho},\Rh_i])\right\}.
\eea
For the spatial moments of inertia we have
\bea
\label{aa}
a^2 =
 \int d^3x \left\{- \tr (\bx \times \bRh)_1^2  -{1\over 4}
\tr([(\bx \times \bRh)_1 ,\Rh_i][(\bx \times \bRh)_1 ,\Rh_i])\right\}
\eea
and  similar formulae for $b^2$ and $d^2$ with the free index $1$ replaced
by $2$ and $3$ respectively.
The moments of inertia in iso-space  are given by
\bea
\label{AA}
A^2 =
\int d^3x \left\{{1\over 4} \tr ([\tau_1,\Uh]\Uh^{\d}[\tau_1,\Uh]\Uh^{\d})
+{1\over 16} \tr ([[\tau_1,\Uh]\Uh^{\d},\Rh_i][[\tau_1,\Uh]\Uh^{\d},\Rh_i])
\right\}
\eea
and similar formulae for $B^2$ and  $C^2$ with $\tau_1$ replaced by
$\tau_2$ and $\tau_3$ respectively. Finally  the coefficient $e$ of the
cross term is
\bea
\label{e}
e =i \int d^3x \left\{-{1\over 2} \tr ([\tau_3,\Uh]\Uh^{\d} (\bx \times
\bRh)_3)
-{1\over 8}\tr ([[\tau_3,\Uh]\Uh^{\d},\Rh_i][(\bx\times
\bRh)_3,\Rh_i])\right\}.
\eea
In all the above integrals the integrands are invariant under the maps
$O_{03},O_{11},O_{12}$. As a result they can be computed by integrating
only over the octant $x_i \geq 0 $ ($i=1,2,3$) and multiplying the result by 8.
Details of how this is done in practice may be found in the appendix.

When $\dot \bS \neq \b0$  the only additional term in the
kinetic energy that is allowed  by the requirement of invariance under
$O_{03},O_{11}$ and $O_{12}$ is
\bea
D(G)_{ik}D(G)_{jl}\dot S_i \dot S_j
M_{kl}
\eea
where the tensor $M_{kl}$ depends on $\rho$ and can be calculated from
\bea
M_{kl} =  \int d^3 x\left\{- {1\over 2}\tr (\Rh_l \Rh_k) - {1\over
8}\tr([\Rh_l,\Rh_n]
[\Rh_k,\Rh_n])\right\}.
\eea
The reflection symmetries of $\Rh_i$ imply that $M_{lk}$ is diagonal,
so that one needs to calculate three more functions of $\rho$ in order
to  study the dynamics in an arbitrary inertial frame.
In the following we will restrict attention to the centre of mass  frame.
Thus we will only compute the Lagrangian on the  manifold $M_7$ of centred
attractive channel fields.

It is useful to  rewrite the
expression  (\ref{form}) slightly by completing the square:
\bea
\label{formm}
T_{\mbox{\tiny att}} = {1\over 2}\left ( f^2 {\dot \rho}^2 + a^2 \omega_1^2
+
 b^2 \omega_2^2
+ c^2 \omega_3^2  + A^2\Omega_1^2
+B^2 \Omega_2^2 + C^2(\Omega_3 + w\omega_3)^2\right),
\eea
with $c^2 = d^2 -e^2/C^2$ and $w=e/C^2$.
The kinetic energy can be expressed in terms of the metric
\bea
\label{metric}
g = f^2 d\rho^2 +a^2 \sigma_1^2 + b^2 \sigma_2^2
+ c^2 \sigma_3^2  + A^2\Sigma_1^2
+B^2\Sigma_2^2 + C^2(\Sigma_3 + w \sigma_3)^2
\eea
where $\sigma_i$ and $\Sigma_a$ ($i,a=1,2,3$) are the left-invariant
one-forms which,
when evaluated on a tangent vector to a trajectory in $M_{7}$,
give the left-invariant angular
velocities:
\bea
\sigma_i({d \over dt}) = \omega_i \quad \mbox{and} \quad
\Sigma_a({d\over dt}) = \Omega_a.
\eea
Explicitly, one finds the following formulae in terms of the Euler angles
defined earlier
\bea
 \sigma_1 &=&  \sin\psi d\theta - \cos\psi \sin\theta d\phi  \nonumber\\
 \sigma_2 &=&  \cos\psi d\theta + \sin\psi \sin\theta d\phi \nonumber\\
\sigma_3 &=&  d\psi  + \cos\theta d\phi,
\eea
and the corresponding expressions for $\Sigma_a$ with $(\phi,\theta,\psi)$
replaced by $(\alpha,\beta,\gamma)$. One checks that $d\sigma_i =-{1\over
2}\epsilon_{ijk}
\sigma_j\wedge\sigma_k$ and   $d\Sigma_a =-{1\over 2}\epsilon_{abc}
\Sigma_b\wedge\Sigma_c$.
For our discussion we also  require the
left-invariant vector fields $\xi_j$ on $SO(3)^J$  and
$\zeta_b$ on $SO(3)^I$ which are dual to the forms $\sigma_i$ and
$\Sigma_a$,
{\it i.e.}
$\sigma_i(\xi_j) = \delta_{ij}$ and $\Sigma_a(\zeta_b) = \delta_{ab}$.
Explicitly, in terms of Euler angles,
\bea
 \xi_1 &=& \cot\theta \cos \psi {\partial \over \partial \psi}
           + \sin \psi {\partial \over \partial \theta}
            - {\cos \psi \over \sin\theta} {\partial \over \partial \phi}
\nonumber\\
\xi_2 &=& -\cot\theta \sin \psi {\partial \over \partial \psi}
           + \cos \psi {\partial \over \partial \theta}
            + {\sin \psi \over \sin\theta} {\partial \over \partial \phi}
\nonumber \\
 \xi_3 &=& {\partial \over \partial \psi}.
\eea
The corresponding formulae for the  vector fields $\zeta_b$ are obtained
by  replacing  $(\phi,\theta,\psi)$
with $(\alpha,\beta,\gamma)$.

For general $\rho$, the determination of the metric $g$ requires the
computation
of eight functions of $\rho$, but when  $\rho=\rho_0$ the form of the
metric is  constrained by the  invariance  of the field $\hat
U(\rho_0,\bx)$
under
 the $SO(2)$ right action
 (\ref{so2}).
 It follows that we should identify tangent vectors at $\rho_0$  which
are related by the $SO(2)$ right action (\ref{so2id}).
That action can  be used to rotate $\xi_1$ into $\xi_2$ and $\zeta_1$
into $\zeta_2$. Since the kinetic energy must respect this
identification we conclude that, for $\rho = \rho_0$,
$g(\xi_1,\xi_1) =g(\xi_2,\xi_2)$ and
$ g(\zeta_1,\zeta_2)=g(\zeta_2,\zeta_2)$,
which implies
\bea
\label{boltsym1}
\quad a^2(\rho_0)=b^2(\rho_0)
 \quad \mbox{and} \quad
A^2(\rho_0)=B^2(\rho_0).
\eea
Furthermore, the vector field  $\xi_3 + 2\zeta_3$ generating the $SO(2)$
action
is  not defined when $\rho=\rho_0$, and for the metric to be regular
there  we require
that
\bea
\label{boltsym2}
g(\xi_3+2\zeta_3,\xi_3+2\zeta_3) = c^2(\rho_0) + C^2(\rho_0)(2+w(\rho_0))^2
=0.
\eea
Thus, assuming $C(\rho_0)\neq 0$ (equality  would imply that $\zeta_3$ also
has length zero), it follows that
\bea
c(\rho_0) =0 \qquad \mbox{and} \qquad w(\rho_0)= -2.
\eea

For large $\rho$,  $\hat U(\rho,\bx)$
is approximately  of the product form, and  the potential  and kinetic
energy can be
expressed in terms of the mass and the moment of inertia of the
hedgehog field $U_H$ (\ref{hedge}) and the parameter $\lambda$
characterising its profile function.
Such  a calculation was carried out in  \cite{S} for the  product ansatz
constructed from Lorentz-boosted  hedgehog solutions.
We  have  repeated that  calculation
 using the instanton-generated hedgehog field
for the individual Skyrmions
and omitting relativistic corrections. Both of these modifications
actually make the calculations harder: in \cite{S} one  could exploit the
fact that the  hedgehog fields in the product ansatz individually
satisfy the static equations of motion, and it was found that the
relativistic
corrections make  the final answer simpler than it would otherwise be.
Here we   take  the centred product ansatz for the attractive channel
(\ref{prodan})  and act with spatial and iso-spatial rotations to obtain
\bea
C U_H (D(G)^{-1}\bx +\frac 12 R\be_1)
\tau_3  U_H(D(G)^{-1}\bx- \frac 12 R \be_1 )\tau_3 C^{\d}.
\eea
Then we insert this field
into the Skyrme Lagrangian, allowing the $SU(2)$ matrices $C$ and $G$ and
the
separation parameter $R$ to vary with time.

The results can be expressed in terms of a single Skyrmion's mass,
$M=147.2$,
its moment of inertia, $\Lambda=140.1$, and its dipole strength, $p=3.31$.
For the potential energy  we find the asymptotic formula
\bea
\label{Vasy}
V \sim 2M - 1.44 \cd {4\pi p^2 \over R^3},
\eea
and for the moments of inertia
\bea
\label{momasy}
a^2 &\sim & 2 \Lambda \nonumber\\
b^2 &\sim& \frac 12 M R^2 + 2\Lambda - 1.2\cd {4\pi p^2 \over R} \nonumber
\\
d^2 &\sim& \frac12 M R^2 + 2\Lambda -  0.16\cd {4\pi p^2 \over R}
\nonumber \\
A^2 &\sim& 2\Lambda \nonumber \\
B^2 &\sim&  2\Lambda + 2.0\cd  { 4\pi p^2 \over R} \nonumber\\
C^2 &\sim& 2\Lambda -2.0 \cd  { 4 \pi p^2 \over R} \nonumber\\
e   &\sim& -2\Lambda +  0.64 \cd  { 4\pi p^2\over R},
\eea
with corrections of order  $1/R^2$.
It follows that
\bea
\label{cwasy}
  c^2 &\sim& \frac12 M R^2  - 1.04\cd { 4\pi p^2 \over R}\nonumber\\
  w &\sim&  -1  -0.68\cd {4\pi p^2 \over \Lambda R},
\eea
also with corrections of order $1/R^2$.
Apart from the $1/R$ corrections the formulae for the moments of inertia
 can be understood quite
easily in terms of the moments of inertia of the individual Skyrmions
and Steiner's theorem. Finally  the asymptotic form of the radial
metric coefficient is
\bea
\label{fasy}
f^2(\rho) \sim \frac12 M + 2.72\cd {4\pi p^2 \over R^3}.
\eea

As mentioned  in section 3, the parameter $R$ can be identified
 with $\rho$ for large $R$. Thus the above formulae can be compared with
the
  asymptotic form of the numerically calculated moments of inertia.
The agreement is very good  for $\rho > 10$ . In figures  2 -  6 we
plot cubic splines constructed from our numerical values for the  metric
coefficients $a^2,b^2,c^2,A^2,B^2,C^2,w$ and $f^2$;
the values for $\rho >15$   were calculated using
 the asymptotic
expressions (\ref{momasy}) - (\ref{fasy}).
The  numerical values at $\rho=\rho_0$ are
$a^2 =348.5$, $b^2 = 348.4$,
$c^2 =0.0$, $A^2 = B^2 =229.8$, $C^2=130.3$, $w=-2.0$ and
$f^2=199.9$.

\section{Quantisation}

Our next goal is to write down the quantum Hamiltonian in the
attractive channel and in the centre of mass frame. As a   first step we
calculate the classical Hamiltonian from the Lagrangian
$L_{\mbox{\tiny att}} = T_{\mbox{\tiny att}}- V$ in the usual way.
Thus we define  the  momenta conjugate to
the angular velocities $\bom$ and $\bOm$,
\bea
\bL = {\partial L_{\mbox{\tiny att}} \over \partial \bom}
\qquad \mbox{and}\qquad  \bK= {\partial L_{\mbox{\tiny att}} \over \partial
\bOm},
\eea
which are   the body-fixed angular momenta in space and
iso-space. In components,
\bea
\begin{array}{lll}
L_1= a^2\omega_1, & \qquad L_2=b^2 \omega_2, & \qquad
L_3= (c^2+ C^2 w^2)\omega_3  + w C^2 \Omega_3  \\
K_1 =  A^2 \Omega_1, & \qquad K_2 = B^2 \Omega_2 , &
\qquad K_3 = C^2(\Omega_3 + w \omega_3).
\end{array}
\eea
The corresponding  space-fixed  spatial angular momentum  is $\bJ = D(G)
\bL$
and the space-fixed iso-spatial angular
momentum  is
$\bI = D(C) \bK$.
Finally defining $P_{\rho} = \partial L_{\mbox{\tiny att}} / \partial \dot
\rho$
we arrive at the classical Hamiltonian in the centre-of-mass frame
\bea
H_{\mbox{\tiny att}} ={1\over 2}\left( {P_{\rho}^2 \over  f^2}
+{L_1^2\over  a^2} + {L_2^2 \over  b^2} +{(L_3-w K_3)^2 \over  c^2}
+ {K_1^2 \over  A^2} + {K_2^2 \over B^2} + {K_3^2 \over  C^2} \right)
+ V(\rho).
\eea
The conserved quantities are the Hamiltonian itself, $\bJ$ and $\bI$.
It follows that $\bL^2=\bJ^2$ and  $\bK^2=\bI^2$ are also
conserved.

In our  quantisation scheme  the quantum Hamiltonian is
\bea
H = -{\hbar^2 \over 2} \Delta + V.
\eea
Here $\Delta$ is the covariant Laplace operator associated with the
metric $g$ (\ref{metric}). Explicitly, in terms of the vector fields
$\xi_i$
and $\zeta_i$ defined in the
previous section,
\bea
\Delta
= {1\over abcABCf } {\partial \over \partial \rho}\left(
{abcABC \over f} {\partial \over \partial \rho} \right)
+ {\xi_1^2 \over a^2}
+{\xi_2^2 \over b^2}
+{(\xi_3 - w \zeta_3)^2 \over c^2}
+ {\zeta_1^2 \over A^2}
+{\zeta_2^2 \over B^2}
+{\zeta_3^2 \over C^2}.
\eea
Physically, $-i\hbar \xi_i$ and $- i \hbar \zeta_i$ are
the components of the operators for the
angular momenta  $\bL$ and   $\bK$ respectively.

For the rest of this paper we will be  concerned with the stationary
Schr\"odinger equation
\bea
\label{schr}
H\Psi = E \Psi.
\eea
The wavefunction $\Psi$ is a
section of a (possibly trivial) -  complex line bundle over
$M_{7}$. The structure of the bundle is fixed
by Finkelstein-Rubinstein constraints, which we summarise as follows
(see also the discussion
 in \cite{BC}).
If a single Skyrmion is quantised as a spin ${1\over 2}$ particle then the
wavefunction for
several Skyrmions   should   pick up a minus sign when  one of the
Skyrmions
is rotated by $2 \pi$. Since a rotation of a single Skyrmion by
$2 \pi$
is a  non-contractible loop in $Q_B$ for any $B\in {\bf Z}$, and since
 moreover all non-contractible loops are homotopic to this one,
wavefunctions are required
to be  sections of a line bundle over $Q_B$  whose holonomy around any
non-contractible loop
in $Q_B$ is $-1$. It follows from the fact that the exchange of two
Skyrmions is a
non-contractible loop that Skyrmions quantised as  half-odd-integer spin
particles are
fermions and Skyrmions quantised as integer spin  particles are bosons.
This result, due to
Finkelstein and Rubinstein \cite{FR}, is
an example of a topological spin-statistics theorem.

In the section 4 we showed that  $M_7$ has the same zeroth and first
 homotopy groups
as $Q_2$. Thus, since  we quantise a single  Skyrmion as a spin $ {1\over
2}$
particle, $\Psi$ should be a  section
 of a non-trivial bundle over $M_7$ such that its holonomy around a
 non-contractible loop in $M_7$ is $-1$.
It is sufficient to impose this for one such loop, and we will use
the loop $\Gamma_{11}$, which, for well-separated Skyrmions,  is the
rotation
of one Skyrmion by $2 \pi$.
The simplest way to implement
the Finkelstein-Rubinstein constraints is to think of $\Psi$
as
a function on
 the double cover $\tilde M_7$  of $M_7$ and  to   impose the  equivariance
condition
\bea
\label{11}
\Psi\circ O_{11} = -\Psi.
\eea
Points related by $O_{03}$ are still identified in $\tilde M_7$, so
$\Psi$   must obey
\bea
\label{03}
\Psi\circ O_{03} = \Psi.
\eea
Then, since $O_{12}=O_{11}O_{03}$ it follows  that
\bea
\label{12}
\Psi\circ O_{12} =-\Psi.
\eea
In accordance with our interpretation of $O_{12}$ this last condition shows
that the two Skyrmions are fermions, {\it i.e.} that the wavefunction
is odd under their exchange. The important point here is that $M_7$
captures
enough of the topology of $Q_2$ for the topological spin-statistics theorem
to hold.

Exploiting  the symmetry of the Hamiltonian we separate variables
by introducing Wigner functions $D^j_{sm}(\phi,\theta,\psi)$ and
 $D^i_{tn}(\alpha, \beta, \gamma)$ on  $SO(3)^J$
and  $SO(3)^I$ respectively, following the conventions of \cite{LL}.
The former   are simultaneous eigenfunctions of
the square of the total spin operator $\bL^2 = \bJ^2 = -\hbar^2
(\xi_1^2 + \xi_2^2 + \xi_3^2)$, the third component
of the space-fixed spin operator $J_3= -i \hbar  \partial /\partial \phi$
and the third component of the
body-fixed spin operator $L_3 = -i\hbar  \partial /\partial \psi$:
\bea
\bL^2 D^j_{sm} = j(j+1)\hbar^2 D^j_{sm}, \qquad J_3 D^j_{sm} = m\hbar
D^j_{sm},
\qquad L_3 D^j_{sm} = s\hbar D^j_{sm}.
\eea
Similarly, $D^i_{tn}(\alpha,\beta,\gamma)$
is an  eigenfunction of
the square of the total iso-spin operator $\bK^2=\bI^2=-\hbar^2
(\zeta_1^2 + \zeta_2^2 +\zeta_3^2)$, the third component
of the space-fixed iso-spin operator $I_3= -i\hbar  \partial /\partial
\alpha$
and the third component of the
body-fixed iso-spin operator $K_3 = -i\hbar  \partial /\partial \gamma$:
\bea
\bK^2 D^i_{tn} = i(i+1)\hbar D^i_{tn}, \qquad I_3 D^i_{tn} = n\hbar D^i_{tn},
\qquad K_3 D^i_{tn} = t\hbar D^i_{tn}.
\eea
The Wigner functions are normalised so that
\bea
\int |D^j_{sm}|^2 \sin \theta \, d\phi\, d\psi \,d\theta ={8\pi^2 \over
2j+1}
\qquad \mbox{and} \qquad \int |D^i_{tn}|^2 \sin \beta \,d\alpha\, d\gamma
\,d\beta
= {8\pi^2 \over 2i+1}.
\eea

Since both spatial and iso-spatial
rotations  by $2 \pi$ are  contractible in $M_7$   we must use Wigner
functions  which are even under such rotations and hence $i$ and $j$ are
positive
integers and $n,t,m,s$ are integers in the  ranges $-i\leq n,t \leq i$ and
$-j\leq m,s \leq j$.
Wigner functions of integer spin transform under the discrete maps
(\ref{03}) and
(\ref{11}) as follows
\bea
\label{cst}
D^j_{sm}\circ O_{03}& = &(-1)^{s}D^j_{sm} \nonumber\\
D^i_{tn}D^j_{sm}\circ O_{11} & = &(-1)^{i+j}D^i_{-t,n}D^j_{-s,m}.
\eea

When looking for bound states of the Schr\"odinger equation (\ref{schr}) we
can fix the quantum numbers $j$ (total spin), $i$ (total iso-spin),
$m$ (the
third
component of the  space-fixed spin) and $n$ (the third
component of the  space-fixed  iso-spin). We are  interested in
two nucleon bound states, so $i=0$ or $i=1$. The
Wigner function for $i=j=0$ (which is  constant) does not satisfy
(\ref{cst}).
For the quantum numbers of the deuteron,  $(i,j)=(0,1)$,
there is exactly one state (we do not count Wigner functions  differing
in $m$ or $n$ as different states), which we write as
\bea
\label{01}
\Psi = \sqrt{3\over 8 \pi^2} D^1_{0m}(\phi,\theta,\psi) u(\rho).
\eea
Inserting  $\Psi$ into (\ref{schr})
we obtain the following differential equation for $u$:
\bea
\label{deutequ}
{1\over abcABCf } {d \over d \rho} \left(
{abcABC \over f} {d u \over d \rho} \right) +\left({2\over \hbar^2} (E-V)
-{1\over a^2} -{1\over b^2}\right)u =0.
\eea

For   $(i,j)=(1,0)$, the iso-vector states,
 there are two allowed angular states.
The ansatz
\bea
\Psi =\sqrt{3\over 8 \pi^2}D^1_{0n}(\alpha,\beta,\gamma)u(\rho)
\eea
leads to the differential equation
\bea
\label{100}
{1\over abcABCf } {d \over d \rho}\left(
{abcABC \over f} {d u \over d \rho} \right) +\left({2\over \hbar^2}(E-V)
 -{1\over A^2} -{1\over B^2}\right)u =0.
\eea
The second possibility is
\bea
\Psi=\sqrt{3\over 16 \pi^2}
(D^1_{1n} +D^1_{-1,n})(\alpha,\beta,\gamma)u(\rho)
\eea
and leads to
\bea
\label{101}
{1\over abcABCf } {d \over d \rho}\left(
{abcABC \over f} {d u \over d \rho} \right) +\left({2\over \hbar^2}(E-V)
-{1\over A^2} -{w^2\over c^2}-{1\over C^2}\right)u =0.
\eea

To study   more  general bound states and scattering  solutions of
(\ref{schr})
 one can use techniques similar to the ones used in \cite{mopos} to study
the
quantum dynamics of two magnetic monopoles in the moduli space
approximation.
One  should make  the ansatz
\bea
\Psi(\phi,\theta,\psi,\alpha,\beta,\gamma,\rho) = \sum_{jsmitn}
D^i_{tn}(\alpha, \beta, \gamma) D^j_{sm}(\phi,\theta,\psi) u^{ij}_{tn,sm}
(\rho),
\eea
where  the sum  runs over the indices  of the Wigner functions, suitably
restricted by conservation laws and
the constraints (\ref{11}) - (\ref{12}).
For bound states, one may fix $i$, $j$, $n$ and $m$, and one will typically
need to
solve a set of coupled ordinary differential equations for the  functions
$ u^{ij}_{tn,sm} (\rho)$, $-i\leq t\leq i$, $-j\leq s\leq j$.  To obtain
  scattering
solutions one necessarily needs to consider an infinite sum over $j$,
and in principle one needs to solve infinitely many systems of coupled
ordinary
differential equations. However, to compute  scattering cross sections
at low energy it is sufficient to study the systems of equations that arise
for small $j$.

In this paper we will only study bound state problems and restrict
attention to the uncoupled ordinary differential equations (\ref{deutequ}),
(\ref{100}) and (\ref{101}).

\section{Quantum Bound States}

To find the lowest  eigenvalue  of the deuteron  equation (\ref{deutequ})
we need
approximate analytic solutions   near $\rho=\rho_0$ and  for large $\rho$.
To find these it is useful to define  the effective  potential
\bea
 V_{\mbox{ \tiny eff}} = V-2M +{\hbar^2 \over 2}\left({1\over a ^2} +
{1\over
b^2} - {1\over 2\Lambda}\right),
\eea
which  tends to zero as $\rho \rightarrow \infty$. In figure  7  we  plot
both the classical potential $V- 2M$ and the effective potential
$V_{\mbox{\tiny eff}}$. Also defining the shifted energy
\bea
\label{eps}
\varepsilon = E -2M -{\hbar^2 \over 4 \Lambda},
\eea
equation (\ref{deutequ}) becomes
\bea
\label{deut}
{1\over abcABCf } {d \over d \rho} \left(
{abcABC \over f} {d u \over d \rho} \right) +{2\over \hbar^2}
(\varepsilon-V_{\mbox{\tiny eff}}) u  = 0.
\eea
For $\rho$ near $\rho_0$,  we define $h= \rho - \rho_0$ , and the
equation
becomes approximately the Bessel equation
\bea
{1\over h} {d\over dh }\left( h {du\over dh} \right) + {2  f^2(\rho_0)
\over \hbar^2}
\left(\varepsilon - V_{\mbox{\tiny eff}}(\rho_0)\right) u= 0,
 \eea
which has  only one  solution that is regular at the origin, namely
 the Bessel function $J_0$ of zeroth order.
Thus, for small $h$
 \bea
\label{small}
u(\rho_0 + h)\propto J_0 \left( \sqrt{ {2 f^2(\rho_0) \over \hbar^2}
 \left(   \varepsilon - V_{\mbox{\tiny eff}} (\rho_0) \right) }h \right).
\eea
It follows in particular that $u$ is non-zero at $\rho_0$  and  that
it has vanishing derivative there.

For large $\rho$ the equation  (\ref{deut}) becomes
\bea
{1\over \rho}{d^2 (\rho u) \over d\rho^2}   + {M \varepsilon\over \hbar^2}u
=0,
\eea
which has two solutions, one exponentially growing and one
exponentially
decaying.  A normalisable wavefunction has to be asymptotically
proportional to the latter:
\bea
\label{large}
u \propto { e^{-{\sqrt{-M\varepsilon}\over  \hbar}  \rho }\over  \rho}  .
\eea
Imposing the behaviour (\ref{small}) and (\ref{large})
 near $\rho_0$
and for large $\rho$
we have numerically searched  for normalisable eigenfunctions of
(\ref{deut}), using a shooting method; we find
that there is precisely  one,  which   we  denote by $u_d$ and which we
normalise
so that
\bea
\int_{\rho_0}^{\infty}d\rho \,  u^2_d  abcABCf =1.
\eea
 The corresponding  eigenvalue is
$\varepsilon_d= -1.107  $ in geometrical units, or $\varepsilon_d= -6.18$
MeV
in physical units. We also note that  the corresponding eigenvalue
of the original  Schr\"odinger equation  (\ref{schr}) is, according
to (\ref{eps}), $E_d = 2M + \hbar^2/4\Lambda + \varepsilon_d =1658.4$ MeV
in physical units.
In the following, the wavefunction (\ref{01})  with $u_d$ for  $u$ and $m=1$
(`spin up'), namely
\bea
\label{deuteron}
\Psi_d = \sqrt{3\over 8 \pi^2} D^1_{01}(\phi,\theta,\psi) u_d(\rho)
= -\sqrt{3\over 16 \pi^2} \sin\theta e^{i\phi} u_d(\rho)
\eea
will be referred to as the deuteron wavefunction.
In figures 8.a) and 8.b) we show  plots of $u_d$ and of the normalised
 probability density
$  u_d^2(\rho) abcABCf$.

The  translation of the eigenvalue $\varepsilon_d$ of (\ref{deut}) into a
     theoretical prediction for the deuteron binding  energy requires some
 thought. Physically, the deuteron's binding energy is defined  as the
energy required to break a deuteron up into two infinitely separated
nucleons,
one proton and one neutron. However, in this paper  we are only considering
two-Skyrmions in the attractive channel, and therefore
wavefunctions on $M_7$, the
space
of attractive channel fields, can only describe  two nucleons with
correlated
spin and iso-spin. Thus, while the energy of two free nucleons is
$2M + 3\hbar^2/4\Lambda = 1708.2$ MeV,  the eigenvalue $\varepsilon_d$
measures
the binding relative to the energy of infinitely separated two nucleons
in the attractive channel
 and with total iso-spin and spin quantum numbers $(i,j)=(0,1)$, which is
  $2M + \hbar^2/4\Lambda$ or 1664.6 MeV. However, it would be
 naive to simply add the  difference of $-\hbar^2/2\Lambda = -43.6$MeV to
  $\varepsilon_d$  and interpret the result as the true
 binding energy of the
deuteron. Rather, a precise calculation of the
deuteron's binding energy would require solving the Schr\"odinger equation
on the full twelve-dimensional moduli  space $M_{12}$ discussed  in the
introduction.
That is a difficult task, but it may be possible to take into account the
extra degrees of freedom approximately by treating them as small
oscillations
around the attractive channel fields. Then one might be able to estimate
the effect of the extra degrees of freedom on the Schr\"odinger equation
for
the deuteron by including their zero point  energies in the effective
potential
$V_{\mbox{\tiny eff}}$. These  zero point energies depend on $\rho$, but
in a first, crude approximation one may describe the   net effect  by
adding a
constant to $V_{\mbox{\tiny eff}}$.
 From  the above discussion of well-separated nucleons in our model we know
that the constant should be approximately $43.6$ MeV.
It then follows that the binding energy of $-6.18$ MeV, which we calculated
by
restricting attention to attractive channel fields, is also  an
estimate of the binding energy  calculated in a more careful treatment
involving the  larger manifold of collective coordinates $M_{12}$.

It is interesting to convert the  quantum mechanical
probability distribution $|\Psi|^2(\rho,\phi,\theta,\psi)$
of a quantum state  $\Psi$
 on the moduli space into a  probability distribution $p_{\Psi}(\bx)$
on physical space in such  a way that $p_{\Psi}(\bx)d^3x$ can be
interpreted
as the probability  of finding one of the nucleons in the region
$[x_1,x_1 +dx_1]\times [x_2,x_2 + dx_2]\times [x_3,x_3 +dx_3]$.
Strictly speaking such a translation only makes sense for Skyrme fields
consisting of two well-separated Skyrmions.
For such fields  the baryon number density $B^0$  is peaked near the
individual
 Skyrmions' positions  and  may be interpreted as  the `particle density'
of the
  Skyrme field.
When the Skyrmions coalesce
it is no longer meaningful to talk about the individual Skyrmions, but
one can still interpret the baryon number density as the `density of Skyrme
matter'.
To find the probability distribution $p_{\Psi}(\bx)$ of Skyrme matter
for a given
quantum state  $\Psi$ on the moduli space,  one should average the
classical baryon number density over the moduli space, weighted by
$|\Psi|^2$.
 Thus, writing  $\hat B^0(\rho,\bx)$ for the baryon number density
(\ref{bdensity}) of
an attractive channel field  in the standard orientation  and  with
separation parameter $\rho$, we define the
spatial probability distribution $p_d(\bx)$  for the deuteron state as
\bea
p_d(\bx) = {1\over 2}
\int   \hat{B}^0(\rho,D(G)^{-1}\bx)
|\Psi_d|^2 \,\sin \theta \,abcABCf\,d\rho\,  d\theta \,d\phi\, d\psi\, .
\eea
Here the $SO(3)$ matrix $D(G)$ is parametrised in terms of the Euler
angles $(\phi,\theta,\psi)$ according to  (\ref{Euler}).
To evaluate this expression we introduce spherical coordinates
$(r,\Theta,\Phi)$ for $\bx$, so that $\bx =
r(\sin\Theta\cos\Phi,\sin\Theta\sin\Phi,
 \cos\Theta)$
and a further set of  spherical coordinates $(r,\thetat,\phit)$ for
$\tilde \bx = D(G)^{-1}\bx$, {\it i.e.}
$\tilde \bx =r(\sin\thetat\cos\phit,\sin\thetat \sin\phit,\cos\thetat)$.
 Then, by expanding $\hat B^0(\rho,\tilde \bx)$ for fixed $\rho$ and
 $r$  in spherical harmonics $Y_{lm}(\thetat,\phit)$, and  by using
  the  transformation properties of spherical harmonics under rotations,
  one shows  that $p_d$  depends only
  on $r$ and $\Theta$, and that it is given
by

\vbox{
\bea
p_d(r,\Theta) &=&{1\over 8 \pi} \int \hat B^0(\rho,
\tilde \bx) u^2_d(\rho) \,abcABCf\,\sin \thetat  \, d\rho\, d\thetat \,
d\phit \nonumber \\
&-&
{1\over 32 \pi} (1-3\cos^2\Theta )\int \hat B^0(\rho,
\tilde \bx) u^2_d(\rho) (1- 3\cos^2\thetat)  \,
 abcABCf\, \sin\thetat \,d\rho \, d\thetat\,
d\phit. \nonumber \\
\eea}

In the  conventional description of the deuteron as a bound state of
two point-like nucleons,  the  square of the modulus of the deuteron
wavefunction  also gives the probability of finding one of the nucleons
in a given region of space. It is therefore meaningful to compare
$p_d(r,\Theta)$ with such a probability distribution calculated in
a conventional model. That  comparison is made in figure  9.
The distributions are clearly  very similar.
Thus, although the classical toroidal bound state of two Skyrmions
looks radically different from the conventional picture of the
deuteron (a point often raised  as a criticism of the Skyrme model)
the two approaches lead to remarkably similar spatial probability
distributions
of nuclear matter at the quantum level.

To study the quantum states with $(i,j)=(1,0)$ we need to
consider  the two
radial  equations (\ref{100}) and (\ref{101}).
There is no bound state solution of (\ref{101}), essentially because
 $c^2$ vanishes at $\rho_0$, giving rise to  a strongly repulsive
  centrifugal potential $w^2/c^2$.
 The equation (\ref{100}), however, does have
a bound state solution.
It can be brought into the same  form  as the deuteron equation
(\ref{deut}), but now the
 shifted energy $\varepsilon$ is
\bea
\varepsilon = E -2M -{\hbar^2 \over 2 \Lambda}
\eea
and  the effective potential is
\bea
 V_{\mbox{ \tiny eff}} = V-2M +{\hbar^2 \over 2}\left({1\over A ^2} +
{1\over
B^2} - {1\over \Lambda}\right),
\eea
which  tends to zero as $\rho \rightarrow \infty$.
There is a unique bound state solution which, like the deuteron
wavefunction,
 is   non-vanishing  at $\rho_0$ and
has  zero derivative there. We will not go into the  details here,
but simply note the numerical value of the bound state energy, which is
$\varepsilon_{NN}= -1.74$ in geometrical units, or $\varepsilon_{NN}
 =-9.74$ MeV
in physical units. It follows that
$E_{NN}=1676.7$ MeV.

 In nature, the deuteron is the only bound state
of two nucleons.
However, there is a marginally unbound  state with quantum numbers
  $(i,j)=(1,0)$, the iso-vector $^1S_0$ state. In our model this state
is the one with energy $E_{NN}$.
Our model gives  the right ordering
of energy levels $E_d < E_{NN}$, and an  energy difference $E_{NN}-E_d$
of the right order of magnitude.
Moreover, $E_{NN} + 43.6$ MeV $> 1708.2$ MeV, the energy of two free
nucleons.
Thus if, as argued above,  the effect of including the two
extra degrees of freedom is to push the energy levels up by $43.6$ MeV,
then the $(i,j) =(1,0)$
bound state we found here will not persist when all 12 degrees of
freedom of $M_{12}$ are taken into account.

\section{Electrostatic Properties of the Deuteron}

The first attempt to calculate the Skyrme model's predictions
for the deuteron's electrostatic properties  was made by Braaten and Carson
in
\cite{BC} under the assumption that the deuteron can be described as a
quantum state of
 a toroidal Skyrmion.  We will briefly review  and then use
their basic formulae for the  classical  electrostatic properties
of a Skyrme field. Since we are using a larger number
of collective coordinates than  in \cite{BC}, our
calculation of  quantum mechanical expectation values will, however,
be different.

The starting point of the calculations is the
basic relationship, first  derived in \cite{Witten}, between
the electromagnetic current  $j_{\mu}$, the third component
of the (classical) iso-spin current  $I^3_{\mu}$ and the
baryon number current  $B_{\mu}$ introduced in  (\ref{bcurrent}):
\bea
j_{\mu}  =  \frac12 B_{\mu} + I^3_{\mu}.
\eea
We are  only  interested in the  expectation values of various
classical multipole  tensors  for  quantum states
with zero iso-spin, so we set $I^3_{\mu}$ to zero in  the following and
replace
 $j_{\mu}$ by ${1\over 2} B_{\mu}$.
 It is sufficient to perform the computations of the various tensors
with  the  Skyrme fields in their standard
orientation.
Specifically, we will compute the   theoretical
predictions for the deuteron's electric charge radius, its electric
quadrupole moment and its  magnetic dipole moment. We will first discuss
the general formulae for these quantities and then compile the
numerical results in a table.

First consider  the root mean square (rms) electric
charge radius of an attractive channel field. It is a function
of $\rho$ only and  defined as the square root of
\bea
\label{rms}
 r_{\mbox{\tiny rms}}^2 (\rho) = \frac12 \int d^3 x |\bx|^2
B^0(\rho,\bx).
\eea
The theoretical prediction for the deuteron's  rms
electric charge radius  $r_c$ is the square root of
\bea
r_c^2= \langle \Psi_d|r_{\mbox{\tiny  rms}}^2|\Psi_d\rangle =
\int_{\rho_0}^{\infty}
d\rho \, r_{\mbox{\tiny rms}}^2  u_d^2 abcABCf.
\eea

Next,
the classical electric quadrupole tensor for  an attractive channel
field in standard orientation also depends only on  $\rho$
 and is given
by
\bea
\label{Q}
\hat Q_{ij}(\rho) = \frac12\int d^3 x (3 x_ix_j
-|\bx|^2\delta_{ij})\hat B^0(\rho,\bx).
\eea
It follows from the discrete symmetries of the attractive channel fields
that $\hat Q_{ij}$ is diagonal. It is also traceless, so it contains
only 2 independent components ($\hat Q_{11}$ and $\hat Q_{22}$, say).
Moreover, at $\rho_0$,  $\hat Q_{11}= \hat Q_{22}$.
The quadrupole tensor for a field in general orientation $G$ is given by
\bea
Q_{lm} = D(G)_{li} D(G)_{mj} \hat Q_{ij}
\eea
and the deuteron's quadrupole moment is defined as
\bea
Q = \langle \Psi_d| Q_{33}|\Psi_d\rangle .
\eea
Performing the angular integration one finds
\bea
Q =  \frac15\int_{\rho_0}^{\infty} d\rho \, (\hat Q_{11} + \hat Q_{22})
u^2_d abcABCf.
\eea

Finally, the iso-scalar ($I^3_{\mu}=0$) part of the magnetic dipole moment
is defined as
\bea
\mu_i(\rho) = \frac14 \int \epsilon_{ijk}x_j B_k(\rho,\bx)d^3x.
\eea
For fields in the standard orientation
it can be written
\bea
\hat \mu_i = \hat M_{ia} \Omega_a + \hat m_{ik} \omega_k,
\eea
where
\bea
\hat M_{ia} & = &{1\over 32 \pi^2} \int d^3x x_j\tr([\hat R_i,\hat R_j]
[-{i\over 2}\tau_a,\hat U]\hat U^{\d})
\nonumber \\
\hat m_{ik} &=&-{1\over 32 \pi^2}\int d^3 x \epsilon_{kmn} x_j x_m
\tr([\hat R_i,\hat R_j]\hat R_n).
\eea
It follows from the discrete symmetries of the attractive channel fields
that both $\hat M_{ia}$ and $\hat m_{il}$ are symmetric, and that the only
non-zero
component of $\hat M_{ia}$ is  $\hat M_{33}$.
In fact, we shall see shortly that
   we only need $\hat m_{11}$ and $\hat m_{22}$ when computing the
deuteron's
   magnetic dipole moment.
Explicitly one finds
\bea
\label{m}
\hat m_{11}(\rho) = \frac14 \int d^3x (x_2^2 + x_3^2) \hat B^{0}(\rho,\bx)
\nonumber \\
\hat  m_{22}(\rho) = \frac14 \int d^3x (x_1^2 + x_2^2) \hat B^{0}(\rho,\bx);
\eea
 at $\rho_0$ these
two quantities are equal.
To compute the  deuteron's magnetic moment, defined via
\bea
\mu= \langle \Psi_d|\mu_3|\Psi_d\rangle,
\eea
we require the magnetic moments of  the  attractive channel fields
in arbitrary orientation $G$, which are given by
\bea
\mu_l(\rho) = D(G)_{li} \hat \mu_i(\rho).
\eea
Then, in order to calculate the expectation value in the quantum state
of the deuteron, we should replace
$\omega_1,\omega_2$ and $\omega_3$ by the operators $L_1/a^2, L_2/b^2$ and
$L_3/c^2$, and  similarly $\Omega_1,\Omega_2$ and $\Omega_3$ by the
operators
$K_1/ A^2, K_2/ B^2$ and $K_3/C^2$.
Since the components of the body-fixed angular momentum
 operator $\bL$ do not commute
with the entries of the matrix  $D(G)$ there is potentially an operator
ordering ambiguity in evaluating  matrix elements of the operator
$\mu_l$, but in the calculation of the expectation value in the
deuteron state this ambiguity does not arise. Here we find
\bea
\mu = {\hbar \over 2 }\int_{\rho_0}^{\infty}d\rho \, ({\hat m_{11}\over
a^2} +
{\hat m_{22} \over b^2})
 u^2_d abcABCf .
\eea

 In nature, the magnetic dipole moment of the  deuteron is almost
exactly equal to the sum of the magnetic dipole moments
$\mu_p$ and $\mu_n$ of the proton
and the neutron. In  conventional  models of the deuteron as a bound
state of a proton and neutron  the discrepancy can be related to the
d-wave contribution in the deuteron wavefunction
\cite{BW}. It is instructive
for us to carry out a similar comparison, but here we should compare our
result
 for the deuteron's magnetic dipole moment
with the sum of the proton's and neutron's magnetic dipole moments
as calculated in the Skyrme model with the appropriate parameters.

We list our numerical results in the table below. We have followed
the  practice   in nuclear physics
of measuring magnetic dipole moments in
units of the nuclear magneton (nm) $\hbar /2 M_N$, where $M_N =939$ MeV is
the
physical nucleon mass. For comparison we also list the results obtained by
Braaten and
Carson in their treatment of the deuteron as a toroidal Skyrmion.
The experimental values are taken from \cite{Ericson}.
In the first row  we compare values for the deuteron binding energy.
Under Braaten \& Carson we quote the value which Braaten and Carson
calculated  in \cite{BC} by computing the difference between the energy
of their deuteron state and the energy of two free nucleons.
Braaten  and Carson did not in fact attach much meaning to
this particular result of their calculations and pointed out
that one should take into  account vibrational modes of the toroidal
configurations  for a  meaningful calculation of  the deuteron's
binding energy. The comparison in our table shows that the inclusion
of the softest vibrational mode alone - the deformation of  a toroidal
configuration  within the attractive channel  manifold $M_{10}$ -
reduces the prediction for the binding energy by an order of magnitude,
and brings it close to the experimental value.
In the last row, the theoretical predictions for $\mu_p$ and  $\mu_n$ are
calculated in the Skyrme model using the formulae given by Adkins
 {\it et al.}
in   \cite{ANW}.
These formulae depend on  the   profile function for the  hedgehog
field  with unit baryon number and on the choice of the constants
$F_{\pi}$ and $e$. For the latter we use the values assumed throughout this
paper and given in (\ref{units}).
Then, to calculate the value of $\mu_p + \mu_n$ in our model we use the
instanton generated
profile function (\ref{hedgehog});
the value listed under Braaten \& Carson is taken from  the analysis of
Adkins and Nappi  in
\cite{AN}  which is based on
the same values for $F_{\pi}$, $e$ and the pion mass as used by
Braaten and Carson.

\vspace{0.5cm}

\vbox{
\centerline{
\begin{tabular}{|c|c|c|c|}    \hline
&\multicolumn{2}{c|}{}&\\
     & \multicolumn{2}{c|}{Theory}&  Experiment\\
& \multicolumn{2}{c|}{}&\\
\hline
&&&\\
&  this model     & Braaten \& Carson       & \\
&&&\\
\hline
&&&\\
$\varepsilon_d$ [MeV] &  $-6.18$ &   $-158$ &  $-2.225$\\
&&&\\
\hline
&&&\\
$ r_c$ [fermi]  & $2.18$   &   $0.92$ & $2.095$ \\
&&&\\
\hline
&&&\\
$Q$ [fermi$^2$] &  $0.83$ &   $0.082$  &  $0.2859$\\
&&&\\
\hline
&&&\\
$\mu$ [nm]   & $0.55$   &   $0.74$  & $0.8574$ \\
&&&\\
\hline
&&&\\
$\mu_p+\mu_n$ [nm]   & $0.41$   &   $0.55$ (from \cite{AN})  & $0.8797$ \\
&&&\\
\hline
\end{tabular}
}
\vspace{0.4cm}
\centerline{\bf Table 1}
\centerline {\sl  Deuteron   Properties } }

\vspace{0.5cm}

\section{Discussion}

In this paper we  have described the deuteron as a quantum state
of instanton-generated two-Skyrmions in the attractive channel.
This picture seems qualitatively correct and leads to  predictions
for certain  deuteron observables which are also in  reasonable
quantitative agreement with experiment.
However,
our calculations clearly involve  a number of  approximations,
and
 from the point of view  of the  general approach to two-Skyrmion
dynamics  outlined in the introduction, two of
these deserve a more careful discussion.  Firstly, we obtained
our Skyrme fields  by computing  instanton holonomies and not, as suggested
in the introduction, by calculating paths of steepest descent in
the Skyrme model. This means in particular that we cannot take into
account the  physical pion mass. The experience with the static properties
of a single nucleon in the Skyrme model shows that the inclusion of
the pion mass does not affect the results very much. We expect similarly
that the plots 2 - 6 and 7  of the metric coefficients and the potential
would not  be qualitatively different for fields calculated
via paths of steepest descent in the Skyrme model with  a pion mass term.
There is one obvious   difference which   would  occur in the asymptotic
form
of the plots: while the potential and metric coefficients  calculated above
approach their asymptotic values according to a power law, the approach
would be exponentially fast in a model with non-zero pion mass.
This difference would be important if one were to discuss nucleon-nucleon
scattering. However, it does not matter
 much for the calculations presented above because the deuteron wavefunction
  is not very sensitive  to the
asymptotic form of the Hamiltonian.

The other, more serious simplification we made was to consider the
 ten-dimensional
moduli space $M_{10}$ instead of the  twelve-dimensional  space
$M_{12}$ described in the
introduction. The validity of this approximation is  much harder to assess,
but the following comments may elucidate its physical meaning.
It has been known for a long time (see e.g. \cite{JJP})
that, for zero pion mass,  the potential energy for the interaction
of  two well-separated Skyrmions with arbitrary individual positions and
 orientations is, up to a positive  constant of proportionality,
\bea
\label{asy}
 -(1-\cos \chi) {1-3 (\bn\cd\hat \bR)^2 \over R^3},
\eea
 where $\bR$ is the relative position vector, $\hat \bR = \bR/R$
 and $(\bn,\chi)$  are the axis and rotation angle  for the
$SO(3)$ matrix describing the relative orientation. In the attractive
channel, the relative orientation is chosen so as to minimise the potential
energy  at  fixed $R$. Thus, $\chi$ is set to $\pi$ and $\bn$ is  required
to be
orthogonal to $\hat \bR$. This eliminates two degrees of freedom, but it
still allows for rotations of $\bn$ in the plane orthogonal to $\hat\bR$.

The
potential (\ref{asy}) is the classical analogue, in the Skyrme model,
of the one pion exchange tensor potential in conventional nuclear physics
\cite{JJP}.
To calculate nuclear forces from it one should  compute its expectation
value in  states which are tensor products of the free nucleon wave
functions
given in \cite{ANW}. In particular,
 the expectation value of (\ref{asy}) in the triplet  state with total
  spin 1
  and total isospin 0 (the quantum
numbers of the deuteron) is a $3\times 3$ matrix which  can be expressed in
terms of the total spin operator $\bs$. It is,   up to an overall
positive constant of proportionality
\bea
V_{\mbox{\tiny T}} = {2 -3 (\bs \cd \hat \bR)^2 \over R^3}.
\eea
At fixed $R$, this expression
 is  minimal  when   the  total spin is parallel to the separation vector
$\bR$. The assumption of the attractive channel approximation is
that the  relative orientation is always such that the potential is
minimal  at given $R$.
 Thus, working with attractive channel
fields   amounts, in the language of nuclear physics, to assuming that the
total nucleon spin is always aligned with the relative separation
vector or, equivalently, that the torque resulting from the tensor force
is infinitely strong.

In conventional discussions of the deuteron and its properties
the tensor force is responsible for the existence of a  d-wave contribution
to
 the deuteron wavefunction
The d-wave probability  in turn is linked to physical observables. In
 the absence of a d-wave  the deuteron's  electric quadrupole moment
would be zero and
the deuteron's magnetic dipole moment  would equal the sum of
the proton's and the neutron's magnetic dipole moments. Both the   size
of  the quadrupole moment $Q$ and the difference $\mu-(\mu_p +\mu_n)$
are therefore direct measures
of the d-wave probability, which in turn indicates the strength of the
tensor potential.
Thus, since  the   truncation
of $M_{12}$ to  the space of attractive channel fields
systematically
overestimates the strength of the tensor force, it is not surprising that
our  theoretical
predictions for $Q$ and $\mu-(\mu_p+\mu_n)$  are rather large.
Quantising the extra two degrees of freedom included in $M_{12}$, at least
approximately,
may well bring the predictions much closer to the experimental values.

\vspace{2cm}

\noindent{\bf Acknowledgements}

\noindent  RAL holds a Smith Institute Research Fellowship at St. Catherine's
College, Oxford, and thanks Smith System Engineering Limited for
their support. BJS acknowledges an SERC Research Assistantship and
is grateful for the hospitality of the Isaac Newton Institute, Cambridge,
where  part of
this work was carried out. NSM would like to thank Gary Gibbons for
discussions about this project  at an earlier stage.

\noindent This  work has been assisted by the award of research grant GR/H67652
under the SERC Computational Science Initiative.

\vspace{1cm}

\appendix
\section{Numerical Methods}

In this appendix we give a detailed and in parts technical  account of our
method
for computing the
  potential $V$ in  the attractive channel, the metric
$g$ derived from the kinetic energy $\Tatt$, and the quantities
required in the calculation of static deuteron properties.
  All these depend  only
 on the Skyrmion separation $\rho$, and
 in equations (\ref{PE}), (\ref{ff})-(\ref{e})  and
  (\ref{rms})-(\ref{m}) we have shown
how to write
 them as
 integrals over physical
space of various combinations of the Skyrme fields $\Uh(\rho,\bx)$ and
their currents. The metric $g$ has  eight  non-vanishing components,
and it turns out that the deuteron's  static properties can be expressed
in terms of three independent moments of the baryon number density
$\hat B^0$. Including the potential,
 there are  thus twelve functions of $\rho$ to be computed.

The computational method is similar to that used in \cite{LM} to find
classical
bound states of three and four Skyrmions, although several additions and
refinements are necessary to carry the programme through successfully.  The
following paragraphs concentrate on these new features; \cite{LM}
should be consulted for the remaining details.  There are two tasks:
firstly the solution of the  holonomy equation (\ref{holeq}) to construct
the attractive channel Skyrme fields in standard orientation,
 and then integrations over all space to
obtain the required functions of $\rho$.

Recall that
  the attractive channel
Skyrme fields
are obtained from instantons which
depend on  two essential parameters. In the standard
orientation, for which the relevant Hartshorne data is depicted in
figure 1, these may be taken to be the scale $L$ and the angle
$\vartheta$. Thus, the solutions of the holonomy equation
(\ref{holeq}) for these instantons are  Skyrme fields on ${\bf R}^3$
which also depend on $L$ and $\vartheta$; we denote them by $
\hat U(L,\vartheta,\bx)$. It follows from  the definition of $
\hat U(L,\vartheta,\bx)$ and from  the scaling behaviour
of JNR instantons that
\bea
\label{scaleU}
\hat U(L,\vartheta,\bx) = \hat U(1,\vartheta, {\bx \over L}).
\eea
Hence  the current $\hat \bR_i (L,\vartheta,\bx)
=(\partial_i\hat U(L,\vartheta,\bx) ) U^{\d}(L,\vartheta,\bx)$
satisfies
\bea
\label{scaleR}
\hat \bR_i (L,\vartheta,\bx)={1\over L}\hat \bR_i^L (1,\vartheta,{\bx \over
L}),
\eea
where on the right-hand side the  superscript $L$
indicates that  the differentiation should be carried out
with respect to $\bx/L$.
To obtain the Skyrme fields $\hat U(\rho,\bx)$ used in the main body of
the paper,
the scale $L$ is fixed  for each value of $\vartheta$  at
 the value $L(\vartheta)$  which minimises
the potential energy (\ref{PE}).
Explicitly,
 writing $E_2$ and $E_4$ for the quadratic and the quartic terms in
 the potential energy (\ref{PE}) evaluated on the field
 $\hat U(1,\vartheta,\bx)$, this is
\bea
 L (\vartheta) = \sqrt{E_4 \over E_2}.
\eea
In practice, it is most convenient to compute all fields and currents
at the scale $L=1$ and to find the fields
and currents at the relevant scale $L (\vartheta)$
using the formulae (\ref{scaleU}) and (\ref{scaleR}).

It remains to explain the computation
of the current $\bRh_{\rho}$ (\ref{cur}),
which requires some care. After fixing the scale at $L(\vartheta)$
the formula $\rho = 2 L(\vartheta)(1-\sin\vartheta)$ establishes
a  one-to-one relation between $\rho$ and $\vartheta$. Thus, both
$\vartheta $ and $L$ may be thought of as functions of $\rho$, and
we can write
\bea
\label{radcurrone}
\partial_\rho\Uh(L,\vartheta,\bx) =\Bigl({dL\over d\rho}\Bigr)\partial_L
\Uh(L,\vartheta,\bx)
    + \Bigl({d\vartheta\over
d\rho}\Bigr)\partial_\vartheta\Uh(L,\vartheta,\bx).
\eea
Using the relation (\ref{scaleU}), the derivative of $\Uh$ with respect to
$L$ may be expressed in  terms of the currents $\Rh_i$:
\bea
\label{radcurrtwo}
\partial_L\Uh(L,\vartheta,\bx) \Uh^{\d}(L,\vartheta,\bx)
= -
\sum_i\Bigl({x_i\over L}\Bigr)\,\Rh_i(L,\vartheta,\bx).
\eea
Thus it is the derivative of $\Uh$ with respect to $\vartheta$, or
equivalently the current
 $\Rh_\vartheta=(\partial_\vartheta\Uh)\Uh\sp\d$, that  has to  be calculated
 directly from
the holonomy equations. Finally, the linear
combination on the right-hand side of (\ref{radcurrone})
 is computed after  the scaling factors $L(\vartheta)$ have been calculated;
$dL/d\rho$ and $d\vartheta/d\rho$ may then be found efficiently
by fitting a cubic spline to the values of $L$ at each $\vartheta$.

For further details of the numerical   integration of the holonomy equation
(\ref{holeq}) we refer the reader to \cite{LM}. In particular it is
explained there
 how the currents $\Rh_i$  can be  found directly from a
holonomy equation involving the differentiated instanton data;
the current $\Rh_{\vartheta}$ may
 be treated analogously.
This is preferable to taking numerical derivatives of the Skyrme fields.
Fields and currents are generated on a mesh of $50\times50\times50$ points.
However, because of  the  discrete symmetries (\ref{dissym}), it  is
sufficient to compute the fields and currents on a mesh of
 $25\times25\times25$ points.

Consider now the calculation of the potential, metric and static deuteron
properties from the fields $\Uh(\rho,\bx)$ and the currents
$\Rh_i(\rho,\bx)$ and $\Rh_{\rho}(\rho,\bx)$.
It is convenient to define quantities $u_0$, $\bu$, $\ba_i$ and $\ba_{\rho}$
via
\bea
\label{holorecast}
\Uh &=&u_0 +i\bu\cd\btau
\nonumber\\
\Rh_i &=&i\ba_i\cd\btau
\nonumber\\
\Rh_\rho &=&i\ba_\rho \cd\btau\, ,
\eea
and  to
introduce  the
following  abbreviations:
\bea
\label{newnotation}
\bb_i&=&\epsilon_{ijk}x_j\ba_k
\nonumber\\
\boldf(\balpha)&=&\bu\times\balpha+u_0\balpha
\nonumber\\
\bF(\balpha,\bbeta)&=&\boldf(\balpha)(\bu\cd\bbeta)-\boldf(\bbeta).
   (\bu\cd\balpha).
\eea
Then the  integrand of the potential energy (\ref{PE}) takes the form
\bea
\label{energydensity}
\ba_i\cd\ba_i+\frac12\bigl((\ba_i\cd\ba_i)
  (\ba_j\cd\ba_j)-(\ba_i\cd\ba_j)(\ba_i\cd\ba_j)\bigr)\, ,
\eea
 and the  terms that appear in the
 integrands of the metric coefficients (\ref{ff}) - (\ref{e})  can
 be written as
\bea
\label{bitsandbobs}
\tr ( \bx \times \bRh)_1^2&=&-2\bb_1\cd\bb_1
\nonumber\\
\tr\bigl(\tuu{1}\bigr)^2&=&8(u_2^2+u_3^2)
\nonumber\\
\tr\bigl(\tuu{3}(\bx \times \bRh)_3\bigr)&=&4if_3(\bb_3\times\bu)
\nonumber\\
\tr\Bigl(\bigl[(\bx \times \bRh)_1,\bRh_i\bigr]\bigl[
(\bx \times \bRh)_1,\bRh_i\bigr]\Bigr)&=
&
  -8\bigl((\bb_1\times\ba_1)^2+(x_2^2+x_3^2)(\ba_2\times\ba_3)^2\bigr)
\nonumber\\
\tr\Bigl(\bigl[\tuu{1},\bRh_i\bigr]\bigl[\tuu{1},\bRh_i\bigr]\Bigr)&=&
   32\sum_i\bigl(\ba_i^2(u_2^2+u_3^2)-f_1^2(\bu\times\ba_i)\bigr)
\nonumber\\
\tr\Bigl(\bigl[\tuu{3},\bRh_i\bigr]
\bigl[(\bx \times \bRh)_3,\bRh_i\bigr]\Bigr)&=&
   -16i F_3(\ba_i,\bb_3\times\ba_i),
\eea
together with cyclic permutations.

The  formula (\ref{bdensity}) for the baryon density $\hat B^0(\rho,\bx)$
for fields in standard orientation is now
\bea
\label{baryondensity}
\hat B^0 =-\Bigl({1\over2\pi^2}\Bigr)\,\ba_1 \cd\ba_2
\times\ba_3,
\eea
and the  static deuteron properties
may be calculated from the following  three moments of the baryon density:
\bea
\label{baryonmoment}
I_i(\rho)=\frac12\int d^3x\,\,x_i^2 \hat B^0(\rho,\bx), \qquad i=1,2,3.
\eea
Explicitly one finds
\bea
\label{staticprops}
r^2_{\mbox{\tiny rms}}&=&I_1+I_2+I_3
\nonumber\\
\hat Q_{11}&=&2I_1-I_2-I_3
\nonumber\\
\hat Q_{22}&=&2I_2-I_1-I_3
\nonumber\\
\hat m_{11}&=&\frac12(I_2+I_3)
\nonumber\\
\hat m_{22}&=&\frac12(I_1+I_3).
\eea

All these integrals have to   be computed for several values of $\rho$.
The integration over ${\bf R}^3$  is facilitated by mapping the whole of
${\bf R}^3$ bijectively to  the finite cube
$C=[-1,1]\times[-1,1]\times[-1,1]$.  The mesh used is uniform
on $C$, but the mapping between ${\bf R}^3$ and $C$ changes with the
Skyrmion separation.  In keeping with the earlier discussion, consider
coordinates $(x_1,x_2,x_3)$, with respect to which the Skyrmions are
separated along the  1-axis.  The cube $C$ has coordinates
$(c_1,c_2,c_3)$,
each taking values in $[-1,1]$.
The form of the mapping is motivated by
the observation that at large separations the Skyrmion positions are
approximately $(\pm L_1,0,0)$ and their size is approximately $\sqrt{L_2L}$,
where $L_1$ and $L_2$ are the lengths of the  semi-major and the semi-minor
axes
of the ellipse in the Harthorne data of figure 1.
The aim is to concentrate mesh points around the Skyrmion centres, where
the  integrands make their largest contributions.
Explicitly,
\bea
\label{mappingone}
x_j ={(\kappa\sqrt{L_2L})c_j\over(1-c_j)^2}
\eea
for $j=2,3$, where $\kappa$ is a dimensionless parameter that controls the
degree to which points are concentrated near the Skyrmion centres; all
calculations here had $\kappa=0.4$.
The relation between $c_1$ and $x_1$ comes in two parts:
\bea
\label{mappingtwo}
x_1 =c_1\left( 4L_1(1-c_1)-(\kappa\sqrt{L_2L})(1-2c_1)\right) \mbox{\qquad
for \qquad}c_1\in [0,\frac12)
\eea
and
\bea
\label{mappingthree}
 x_1 ={L_1+\frac14(\kappa\sqrt{L_2L})(c_1-\frac12)\over(1-c_1)^2}
\mbox{\qquad for\qquad}c_1\in[\frac12,1].
\eea
The two pieces fit together smoothly at $c_1=\frac12$.
The expressions for $c_1$ negative are chosen so that $x_1(c_1)$ is an odd
function.

After the mapping, the integrals  are calculated
via tricubic interpolation, as in
\cite{LM}.
The factors $(1-c_i)^2$ in the denominators of (\ref{mappingone})
and (\ref{mappingthree})
ensure
that all the integrands have vanishing  gradients at the boundary of $C$.
(The denominator in \cite{LM} was simply $(1-c_i)$, which is fine for
integrands falling off at least as fast as $r\sp{-6}$ for large $r$;
here, some of the
metric integrals fall off like $r\sp{-4}$, and this requires a different
power if the tails are to be handled correctly.)

We have computed
the integrals (\ref{energydensity}),(\ref{bitsandbobs}),(\ref{baryondensity})
and (\ref{baryonmoment}) for sixteen different values of $\rho$.
The corresponding values for  $\vartheta$ are all the   multiples of
$2\sp\circ$
in the range $2^{\circ} \leq \vartheta\leq 30^{\circ}$,
 and an additional point at $\vartheta=1^{\circ}$, included
to have a better check against the asymptotic expressions (\ref{Vasy}) -
(\ref{cwasy}).  The range
of values of $\rho$ covered in this way is $[1.71,15.3]$ (the size of an
isolated Skyrmion in these units is $\approx 1$).  To compute all
quantities for each
separation takes approximately one hour on a workstation with an R4000
processor.  There is good agreement asymptotically with (\ref{Vasy}) -
(\ref{cwasy}) and the
additional symmetries (\ref{boltsym1}) and (\ref{boltsym2}) when
$\rho=\rho_0$ are
also clearly
present.  The
integrated baryon number density never leaves the range $[2.0003,2.0049]$,
and there is no evidence for regions of negative baryon density.

\noindent{\bf Note  added}: N.R. Walet has recently also computed the
restriction of the Skyrme Lagrangian to instanton-generated Skyrme fields
of  degree two (FAU-T3-94/1,    hep-ph/9410254).
 He considered
three   one-parameter families of such fields, including  the  attractive
channel fields  studied here.

\pagebreak

\parindent 0pt

\centerline{\large  \bf Figure Captions}

\vspace{0.6cm}

{\bf 1}. The Hartshorne data for instanton-generated
two-Skyrmions in the attractive channel. The circle and the  ellipse are shown
in standard orientation; the isosceles triangle with vertices
 $X_1=(0,L,0,0), X_2=(-L\cos\vartheta,-L\sin\vartheta,0,0) $
 and $X_3=(L\cos\vartheta,-L\sin\vartheta,0,0)$ specifies the JNR data used
 in our calculations.

\vspace{0.6cm}

{\bf 2}. Spatial moments of inertia.

\vspace{0.6cm}

{\bf 3}. The spatial moment of inertia $a^2$, shown  on a
smaller scale than in  figure 2.

\vspace{0.6cm}

{\bf 4}. Iso-spatial moments of inertia.

 \vspace{0.6cm}

{\bf 5}. The cross term $w$ which couples  spin and iso-spin.

\vspace{0.6cm}

{\bf 6}. The radial metric coefficient $f^2$.

\vspace{0.6cm}

{\bf 7}.
The potential $V$ (solid  line)
 and the effective potential $V_{\mbox{\tiny eff}}$ (dashed line) which
occurs in the radial Schr\"odinger equation for the deuteron. Both
are plotted in geometrical units.

\vspace{0.6cm}

 {\bf 8.a}) The radial part $u_d$ of the deuteron wavefunction; for this
 plot the normalisation is chosen so that $u(\rho_0)=1$.

 {\bf 8.b}) The  radial probability distribution $u_d^2\, abc ABCf$
 for the deuteron state,  normalised so that
 $\int_{\rho_0}^{\infty} d\rho\,\, u_d^2\, abc ABCf =1$.

\vspace{0.6cm}

{\bf 9.a}) Equally spaced density contours for  the spatial probability
distribution $p_d$ of
  nuclear matter in  the
deuteron state $\Psi_d$. The distribution is axially
 symmetric about the  3-axis, and shown here  in  the $x_1x_3$-plane.
Both $x_1$ and $x_3$ are measured in fermi.

{\bf 9.b})  Density contours for  the spatial probability distribution of the
nucleons calculated
in a conventional potential model of the deuteron, with the nucleons
treated as point-like particles. (From \cite{EW}, with kind permission of
Oxford University Press)

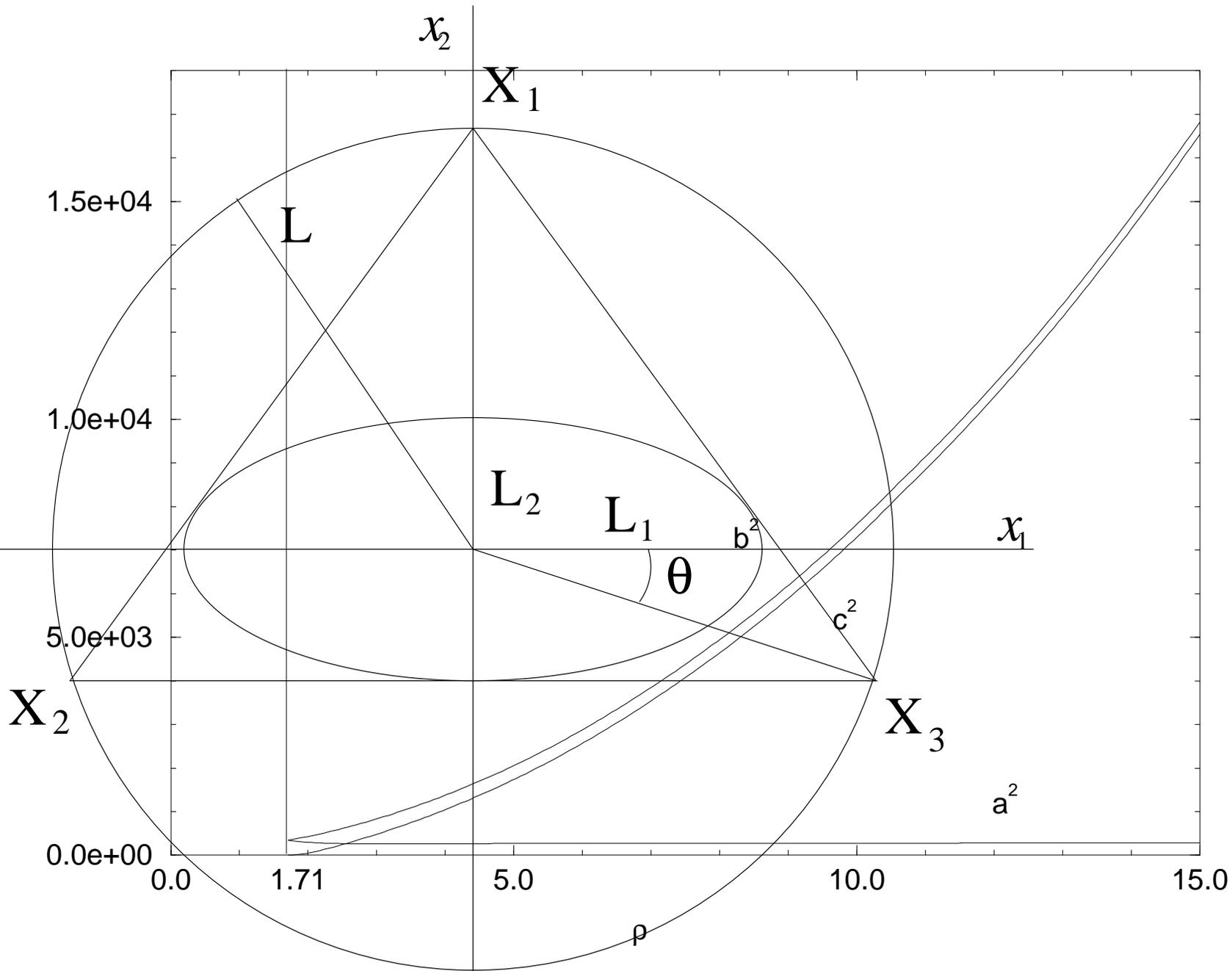
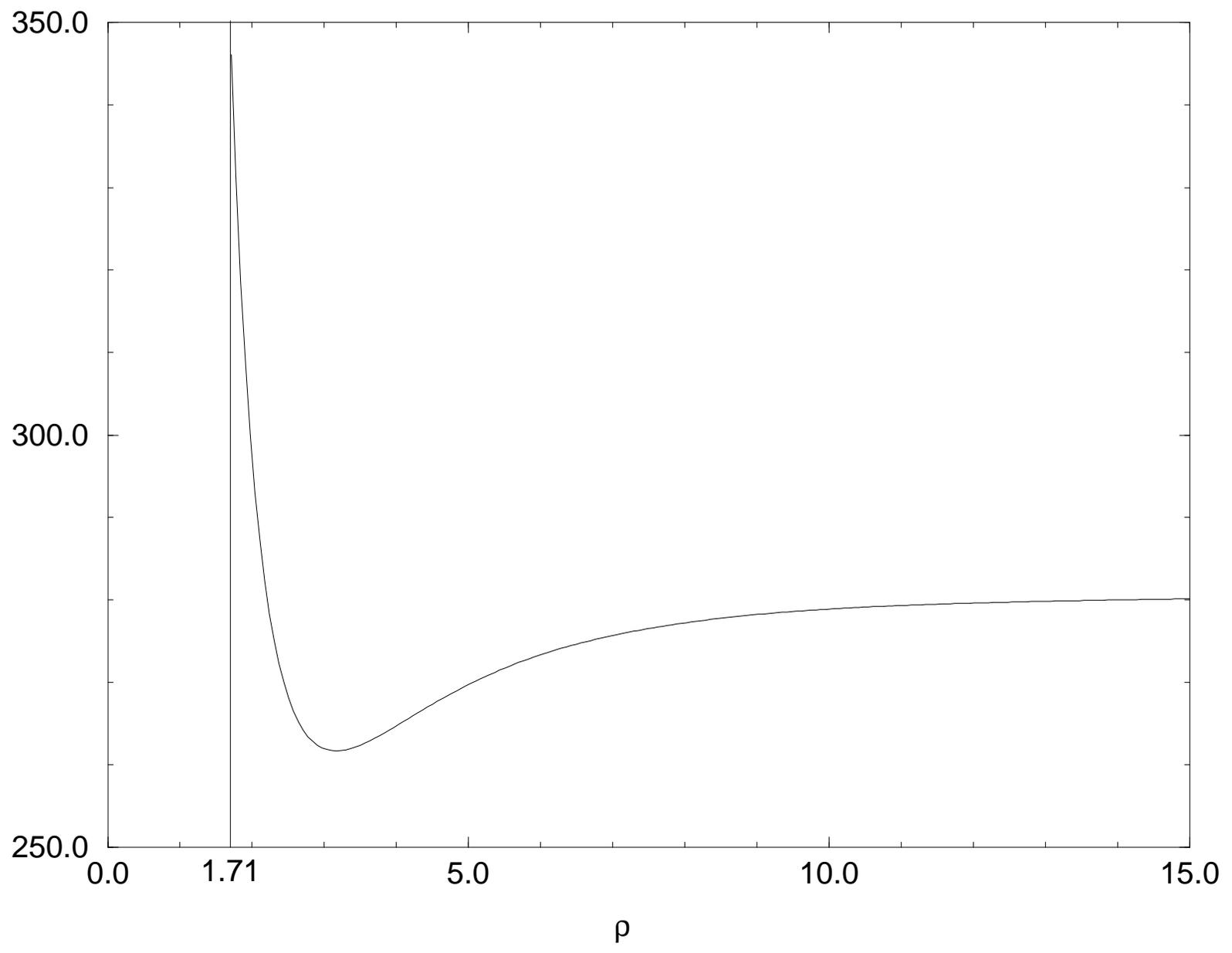
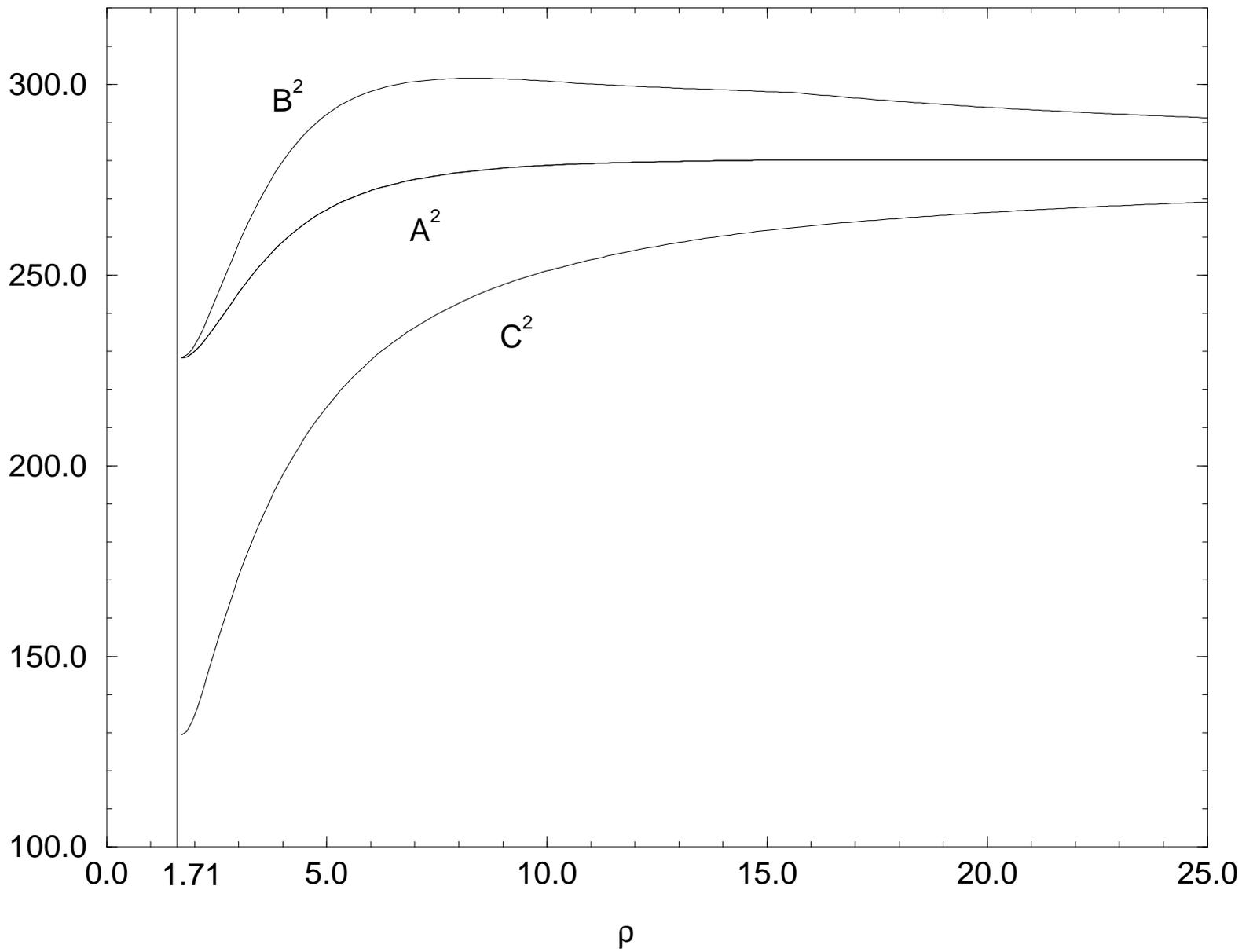
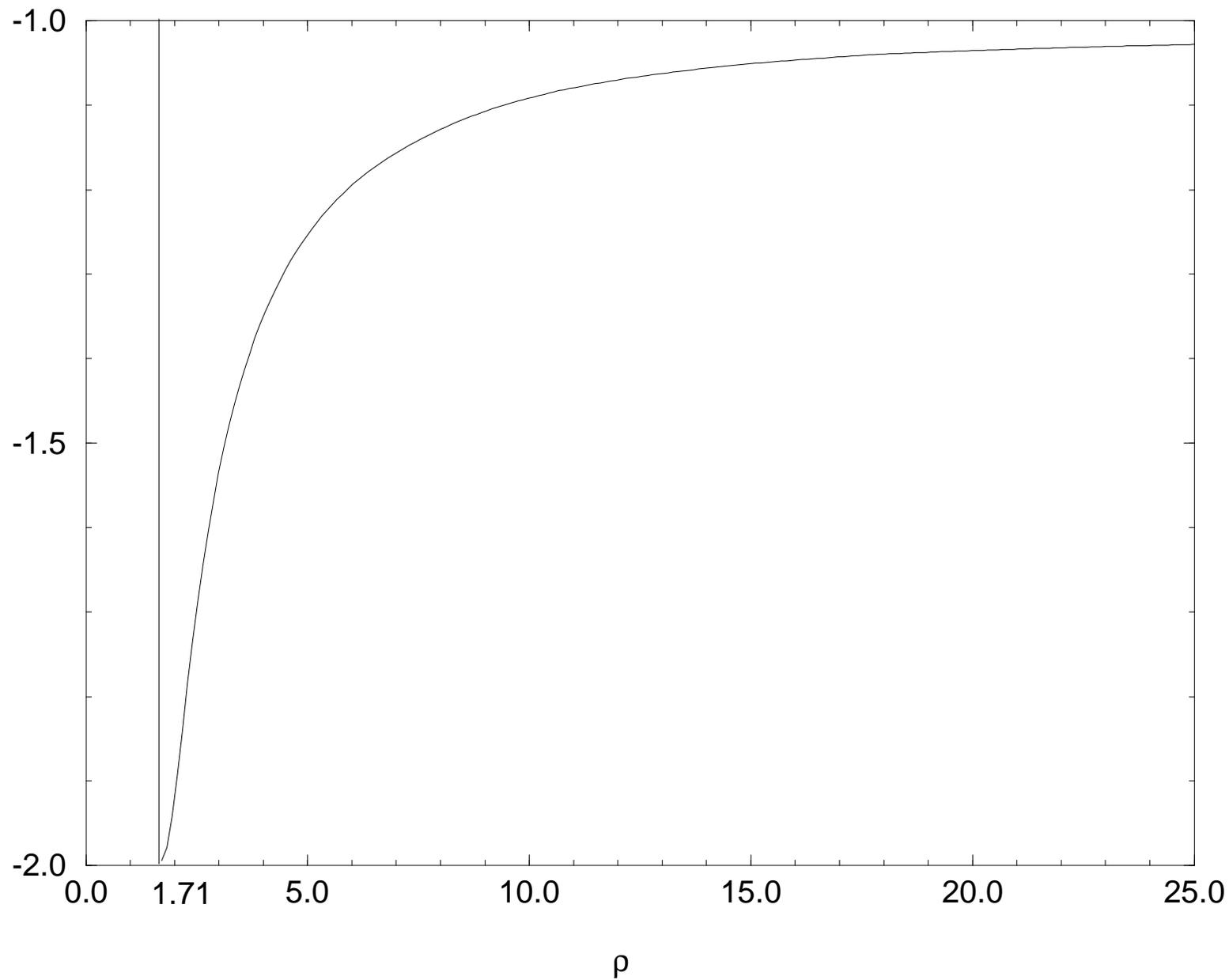
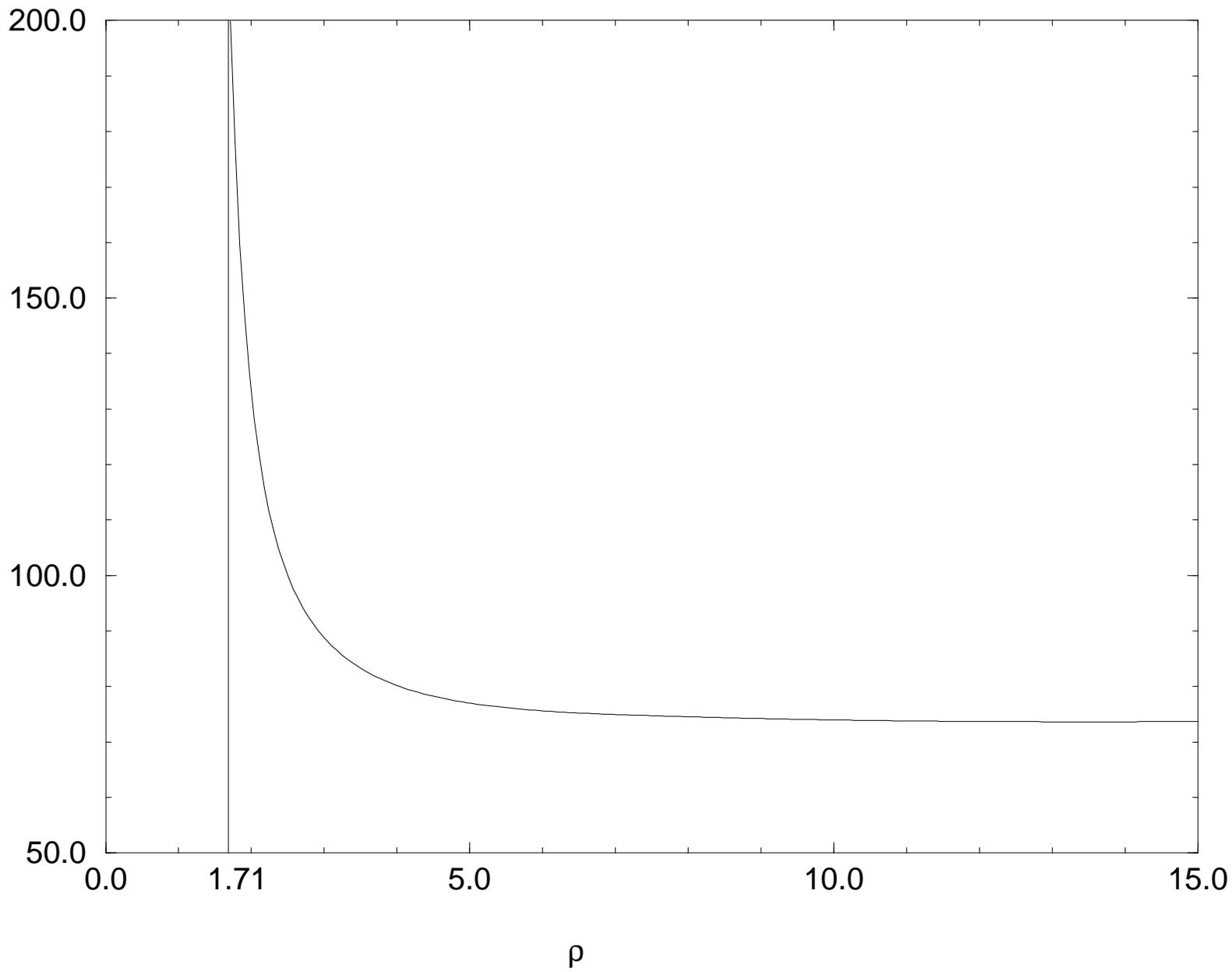
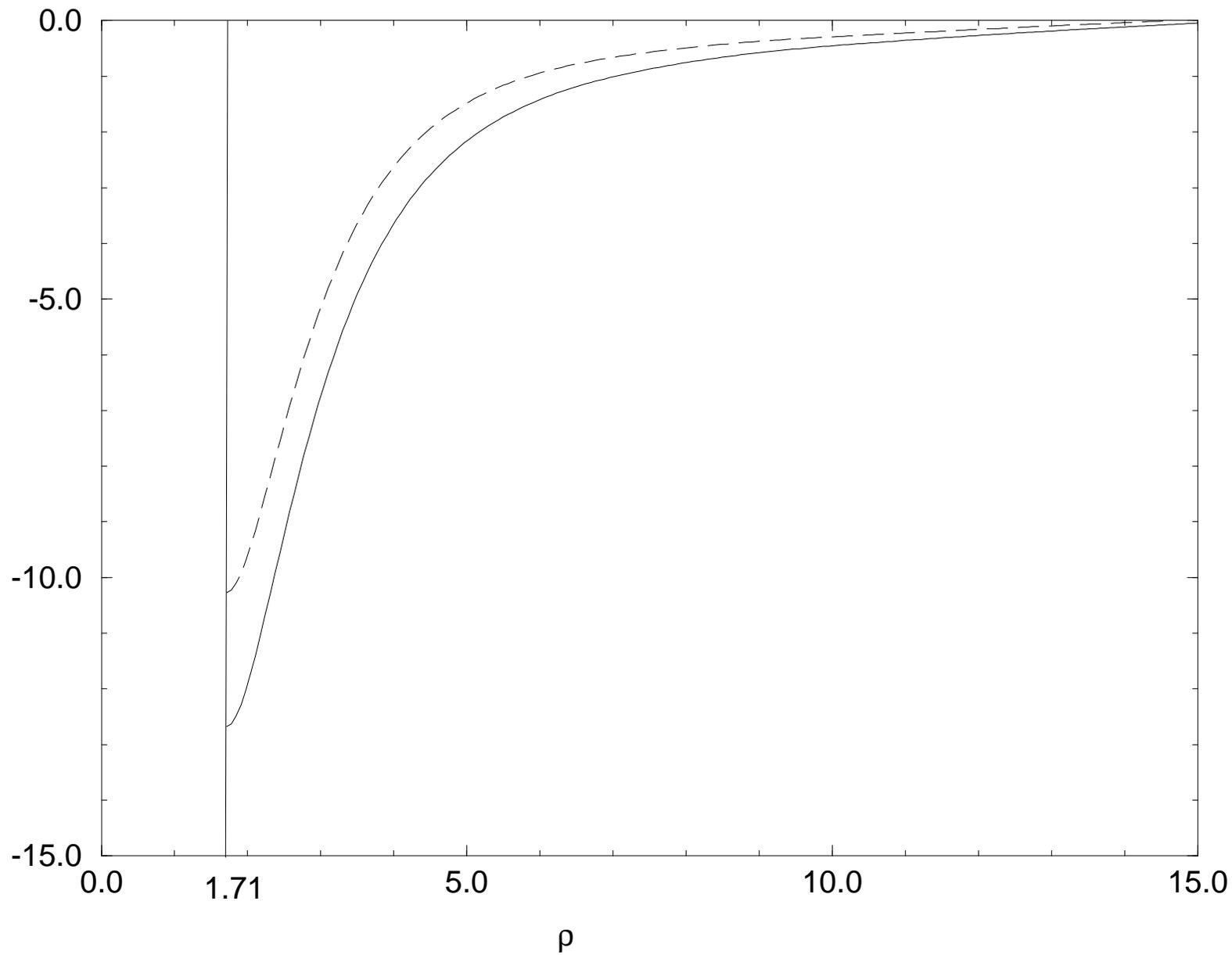
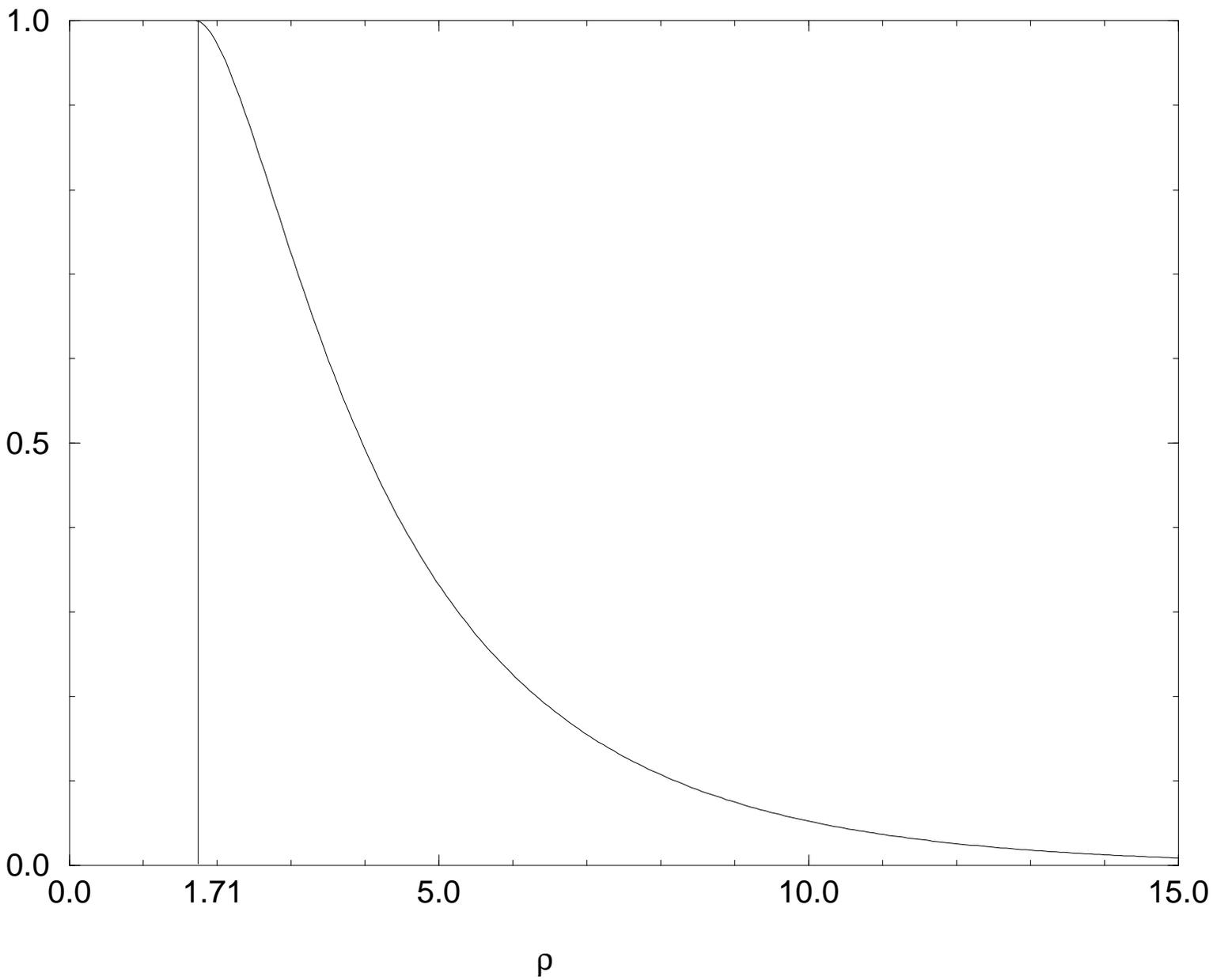
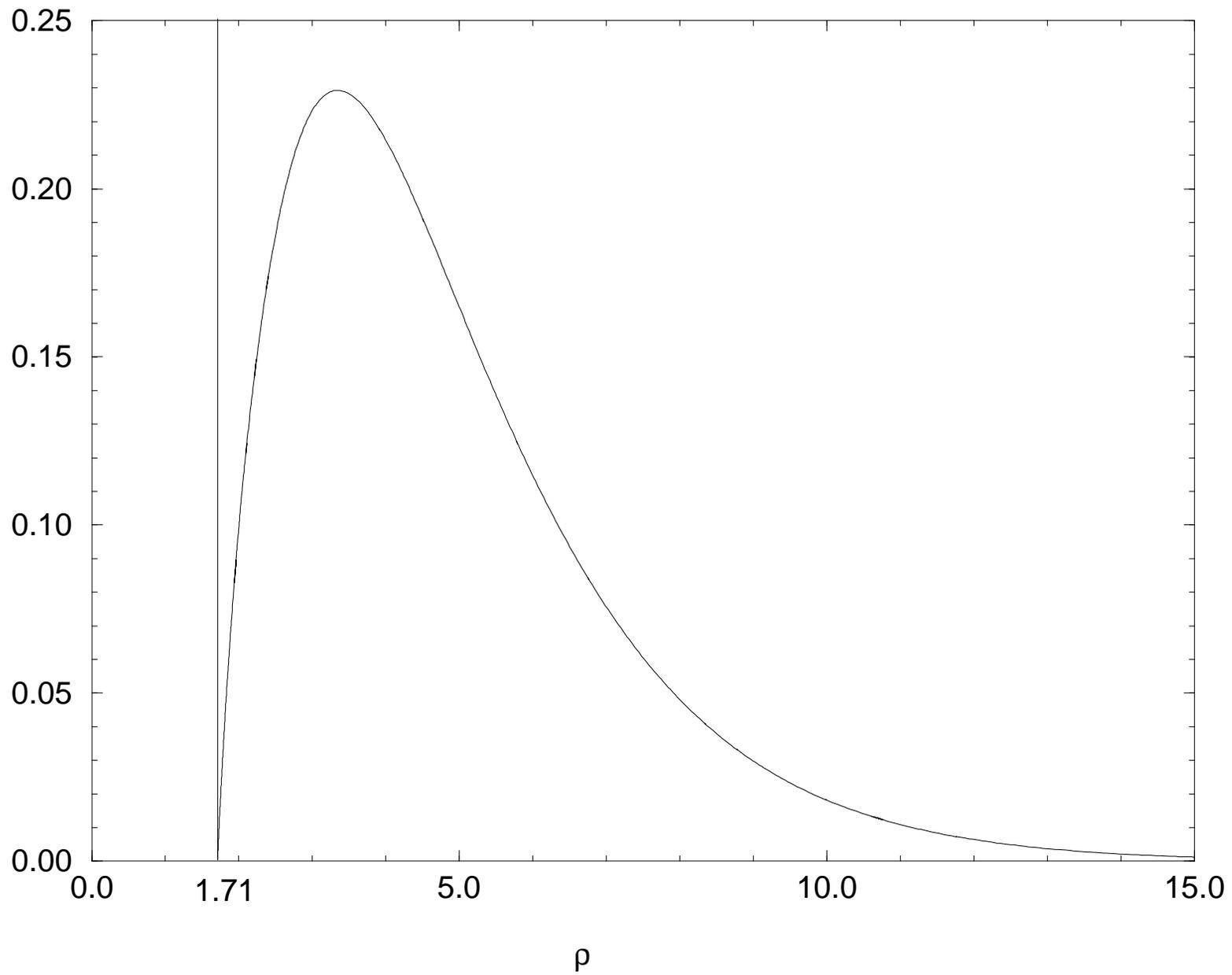
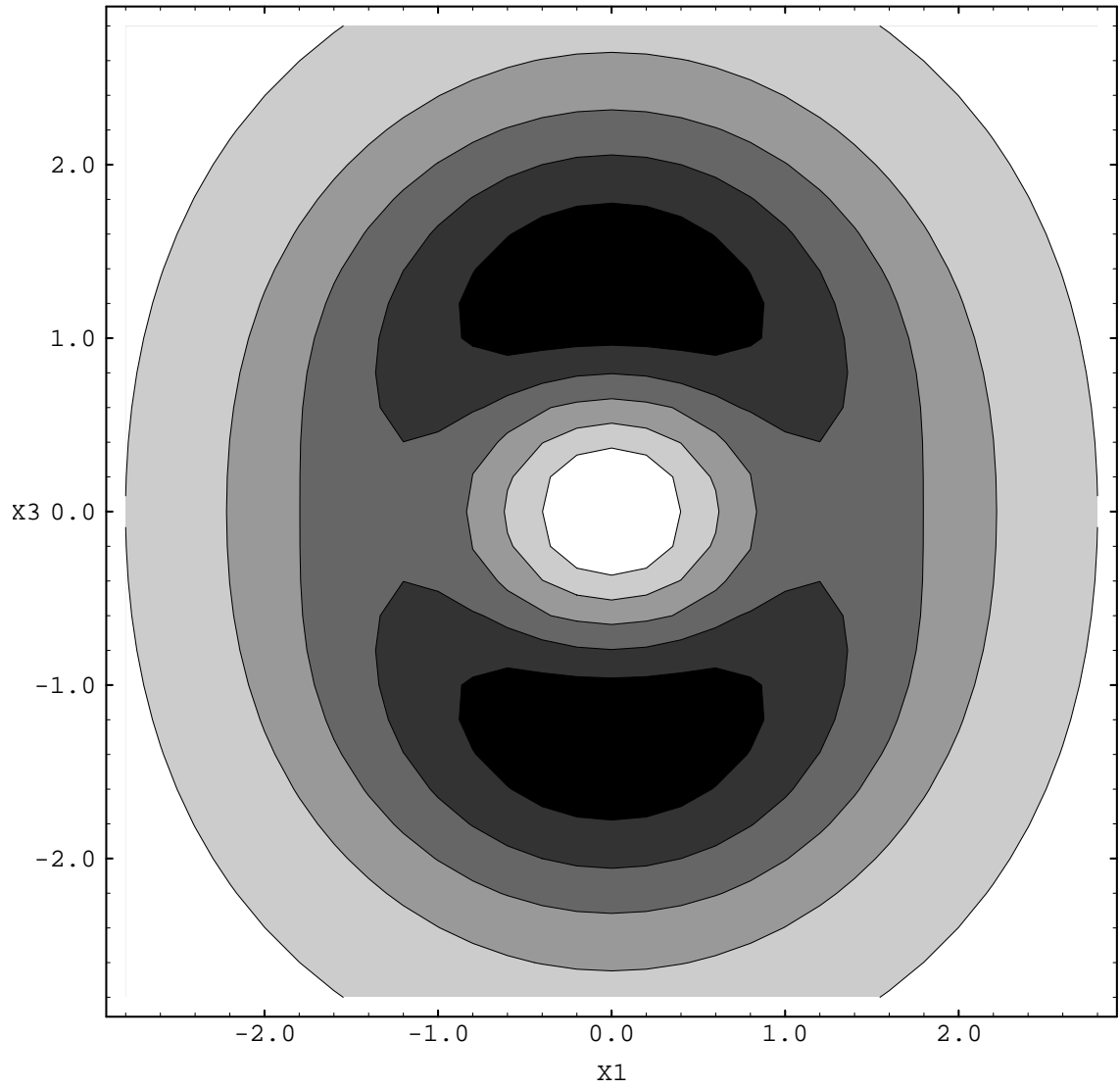
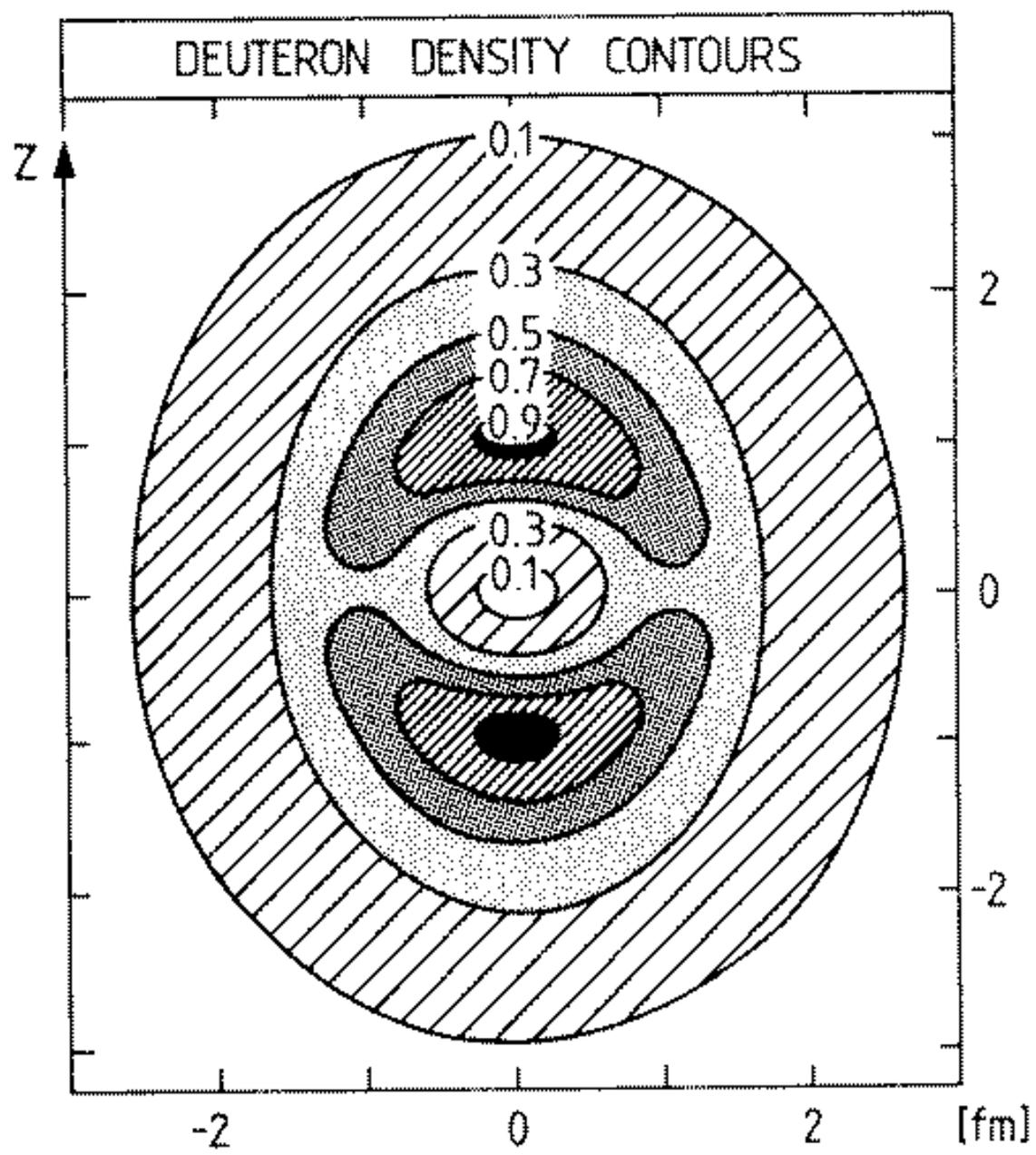
\end{document}